\documentclass[sigconf]{acmart}
\PassOptionsToPackage{table,xcdraw}{xcolor}
\usepackage[table,xcdraw]{xcolor}
\usepackage{hyperref}
\usepackage[nameinlink]{cleveref}
\usepackage{multirow}
\usepackage{listings}
\usepackage{makecell}
\usepackage[frozencache,cachedir=.]{minted}
\usepackage{colortbl}
\usepackage{tabularx}
\usepackage{comment}
\newcommand{\rev}[1]{\textcolor{black}{#1}}

\AtBeginDocument{%
  \providecommand\BibTeX{{%
    \normalfont B\kern-0.5em{\scshape i\kern-0.25em b}\kern-0.8em\TeX}}}

\acmConference[CHI '24]{2024 CHI Conference on Human Factors in Computing Systems}{May 11--16,
  2024}{Honolulu, HI}

\begin{document}

\title{RASSAR: Room Accessibility and Safety Scanning \\ in Augmented Reality}

\author{Xia Su}
\affiliation{%
  \institution{University of Washington, USA}
  \country{}
}
\email{xiasu@cs.washington.edu}

\author{Han Zhang}
\affiliation{%
  \institution{University of Washington, USA}
  \country{}
}
\email{micohan@cs.washington.edu}

\author{Kaiming Cheng}
\affiliation{%
  \institution{University of Washington, USA}
  \country{}
}
\email{kaimingc@cs.washington.edu}

\author{Jaewook Lee}
\affiliation{%
  \institution{University of Washington, USA}
  \country{}
}
\email{jaewook4@cs.washington.edu}

\author{Qiaochu Liu}
\affiliation{%
  \institution{Tsinghua University, China}
  \country{}
}
\email{lqc21@mails.tsinghua.edu.cn}

\author{Wyatt Olson}
\affiliation{%
  \institution{University of Washington, USA}
  \country{}
}
\email{wyatto@uw.edu}

\author{Jon E. Froehlich}
\affiliation{%
  \institution{University of Washington, USA}
  \country{}
}
\email{jonf@cs.washington.edu}

\renewcommand{\shortauthors}{Su et al.}
\renewcommand{\shorttitle}{RASSAR}
\begin{abstract}
The safety and accessibility of our homes is critical to quality of life and evolves as we age, become ill, host guests, or experience life events such as having children. Researchers and health professionals have created assessment instruments such as checklists that enable homeowners and trained experts to identify and mitigate safety and access issues. With advances in computer vision, augmented reality (AR), and mobile sensors, new approaches are now possible. We introduce \textit{RASSAR}, a mobile AR application for semi-automatically \textit{identifying}, \textit{localizing}, and \textit{visualizing} indoor accessibility and safety issues such as an inaccessible table height or unsafe loose rugs using LiDAR and real-time computer vision. \rev{We present findings from three studies: a formative study with 18 participants across five stakeholder groups to inform the design of RASSAR, a technical performance evaluation across ten homes demonstrating state-of-the-art performance, and a user study with six stakeholders. We close with a discussion of future AI-based indoor accessibility assessment tools, RASSAR's extensibility, and key application scenarios.}

\end{abstract}



\begin{teaserfigure}
  \includegraphics[width=\textwidth]{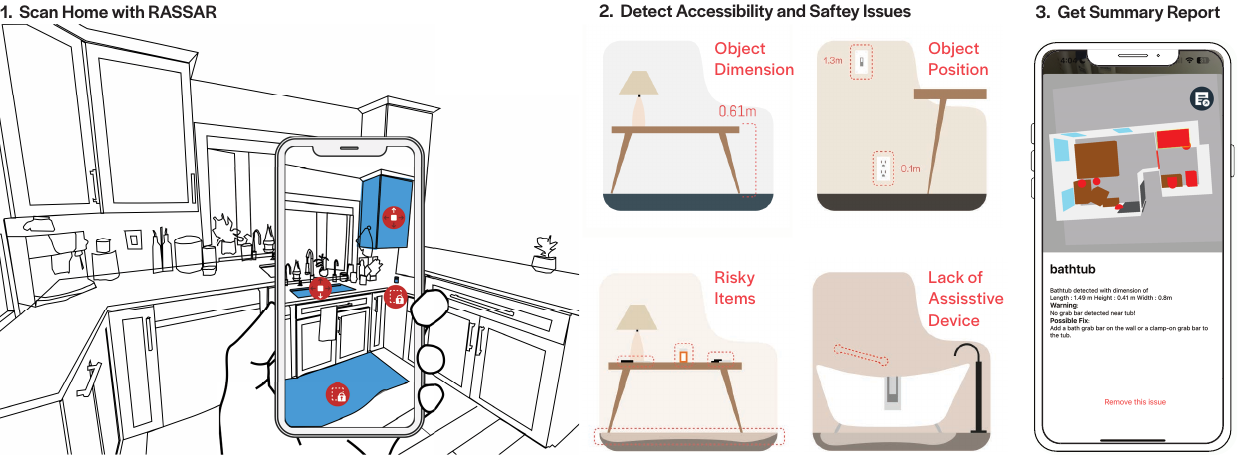}
  \caption{RASSAR is a mobile AR application for semi-automatically \textit{identifying}, \textit{localizing}, and \textit{visualizing} indoor accessibility and safety issues. (1) RASSAR scans home spaces and detects potential issues in real time using LiDAR and computer vision. (2) RASSAR currently supports four classes of issues, including inaccessible \textit{object dimensions} such as a high/low table top or the presence of risky/dangerous items such as scissors. (3) After a scan, RASSAR generates an interactive summary of identified problems with a 3D reconstructed model.}
  \Description[Overview of the RASSAR system]{Left: On the top there is a line says "Scan home with RASSAR". The content is a sketch-style figure showing user holding a phone to scan a kitchen scene. On the phone, there are four red pop-up icons indicating four accessibility or safety issues. The source objects of these issues, i.e. sink, rug, cabinet and medication bottle, are marked in blue. Middle: On the top there is a line says, "Detect accessibility and safety issues". The content is four diagrams showing four categories of accessibility and safety issues that RASSAR can detect. The four categories are: object dimension, whose diagram shows a table with a height mark of 0.61m; object position, whose diagram shows a light switch marked with a height of 1.3m and an electric socket marked with height of 0.1m; risky items, whose diagram shows a scene with four risky items, rug, medicine, knife and scissors; lack of assistive items, whose diagram shows a bathtub and red dashed contour of a grab bar, indicating lack of grab bars. Right: On the top there is a line says, "Get summary report". The content is a UI showing a 3D model of an indoor layout, with some indoor furniture marked in red. The bottom half of the UI shows some elaboration information of a selected furniture, which is a bed. Information such as object dimension and a warning information for bed height is included. At the bottom of the UI is an underlined line shows "Cancel this issue".}
  \label{fig:teaser}
\end{teaserfigure}

\maketitle

\section{Introduction}

Safe and accessible living spaces are a fundamental human right \cite{world_health_organization_who_2018}. Yet, inaccessible housing remains prevalent throughout the world. In the US, for example, 90\% of housing units are inaccessible to people with disabilities \cite{smith_aging_2008,steinfeld_home_1998}. In the UK, 98\% of newly built private homes are inaccessible to wheelchair users \cite{imrie2003housing}. To improve the safety and accessibility of domestic spaces, researchers and health professionals have created pre-formatted checklists that help residents and trained professionals audit and renovate indoor spaces \cite{struckmeyer_home_2021,clemson_home_1997,iwarsson_housing_1999,fange_changes_2005}. For example, the \textit{Home Safety Self-Assessment Tool} (HSSAT) \cite{horowitz_use_2016,tomita_psychometrics_2014} includes a checklist for issues such as uneven flooring, cluttered areas, slippery throw rugs, and inaccessible light switches across nine home areas (\textit{e.g., } kitchens, bathrooms, and bedrooms). Others have explored remote assessment methods via teleconferencing and video cameras \cite{renda_feasibility_2018,romero_development_2017}.

With advances in computer vision (CV), augmented reality (AR), and mobile sensors, new ways to assess indoor accessibility and safety are now possible. For example, emerging smartphones contain built-in \textit{Light Detection and Ranging} (LiDAR) sensors~\cite{apple_apple_2020}, which can reconstruct indoor spaces in real-time with high precision~\cite{diaz_vilarino_3d_2022}, new CV models like \textit{YOLO} \cite{jocher_ultralyticsyolov5_2022} are capable of high recognition rates across object types, and smartphones contain powerful computational units for onboard processing. Leveraging these advances, we created \textit{RASSAR}---\textit{Room Accessibility and Safety Scanning in Augmented Reality}---a custom mobile AR application for semi-automatic\-ally \textit{identifying}, \textit{localizing}, and \textit{visualizing} indoor accessibility and safety issues using LiDAR and real-time computer vision (Figure \ref{fig:teaser}). With RASSAR, a user scans an indoor space with their phone; the tool constructs a real-time parametric model of the 3D scene, attempts to identify and classify known accessibility and safety issues, and visualizes potential problems in AR overlays. Throughout the scan, users can confirm/delete automatic detections or select them to view more information (see \autoref{fig:interface}). After scan completion, RASSAR shows an interactive 3D room reconstruction with a summary of findings (\autoref{fig:teaser} and \autoref{fig:summary}). \rev{Additionally, to assist blind or low vision users, we designed RASSAR to support VoiceOver and provide custom audio assistance about the scanning process (\textit{e.g.,} verbalizations of detected items).}

To develop RASSAR, we conducted a three-stage iterative design process. First, informed by prior work \cite{tomita_psychometrics_2014,horowitz_use_2016,noauthor_ada_2010,housing_fair_1996}, we built a \textit{rapid technical prototype} to demonstrate feasibility and examine initial audit possibilities. Second, using this prototype as a design probe \cite{saha_visualizing_2022,jain_head-mounted_2015,froehlich2012design,hara2016design}, we conducted a \textit{formative study} with 18 participants across five stakeholder groups: wheelchair users, blind and low-vision participants, families with young children, caregivers, and occupational therapists. We showed working videos of RASSAR auditing an apartment, solicited feedback on new interface mockups, and engaged participants in co-brainstorming new features. Study findings reveal common challenges in home accessibility and auditing processes that RASSAR can potentially alleviate, along with suggested improvements to its rubrics and UI. Third, we \textit{built the current RASSAR system} with improved detection performance, user interaction, and interface design and performed two additional studies: a technical performance evaluation (Study 2) and a user study (Study 3) with six participants again drawn from our five stakeholder groups.

As emphasized in literature \cite{struckmeyer_home_2021} and further affirmed by our formative study, home audit tools must be adaptable to address individual needs and differences across community groups. To address this challenge, RASSAR introduces a custom and fully extensible JSON format to encode specific accessibility/safety issues into its audit engine. Currently, RASSAR supports 20 issues across four problem categories: \textit{object dimension} (\textit{e.g.,} table too low), \textit{position} (\textit{e.g.,} light switch too high), \textit{risky items} (\textit{e.g.,} throw rug), and \textit{lack of assistive device} (\textit{e.g.,} no grab bar near toilet). New issues can be added via a text editor with JSON. In the future, such additions could be made via an envisioned authoring tool.


\rev{To evaluate RASSAR's performance and usability as well as its effectiveness in real-world settings, we conducted the aforementioned technical evaluation and user study. In the technical evaluation, the first author visited ten homes of varying sizes and layouts, conducting both a manual baseline audit and a RASSAR scan for comparison. Each RASSAR scan was repeated three times to examine consistency. The user study followed a similar procedure; however, in this case, the participants themselves conducted RASSAR scans independently. We then interviewed participants about their experience to gather qualitative insights. We found an average scan precision/recall of 0.86/0.83 when conducted by the researcher and 0.79/0.73 when scanned by participants. Our findings show that RASSAR is not only usable and useful but also significantly increases efficiency, achieving an auditing speed 3.5x faster than manual auditing. As P4, a wheelchair user, said: ``\textit{RASSAR is easy to use and was pretty accurate in terms of ADA}''.}

In summary, our contributions include: (1) formative findings from five stakeholder groups about their current indoor safety/accessibility audit practices, \rev{their methods for} reconfiguring spaces to fit their needs, and reactions to future indoor auditing tools using advanced sensing; (2) the design and implementation of RASSAR, a novel AR-based tool for semi-automatically detecting safety/accessibility issues in indoor spaces using LiDAR and real-time CV; \rev{(3) findings from a technical performance evaluation and a follow-up user study demonstrating effectiveness and potential applications of RASSAR. As secondary contributions, we introduce an extensible JSON format for specifying indoor safety/access issues and provide a specified object detection model and its training dataset. We have also open sourced RASSAR along with the detection model and its training dataset at \href{https://github.com/makeabilitylab/RASSAR}{\color{blue}{https://github.com/makeabilitylab/RASSAR}}}.

Our work aims to transform how people examine and configure indoor spaces to improve accessibility/safety. We envision RASSAR as a versatile tool to aid builders in considering and validating the safety and accessibility of new construction, residents in planning renovations or updating their homes due to life changes (\textit{e.g.,} illness, birth), rental agencies like Airbnb in vetting and validating the accessibility and safety of rental spaces, and occupational therapists in assisting residents as they comprehensively assess and identify safety/access issues during home visits.

\section{Related Work}
We contextualize our work within the home accessibility and safety auditing literature as well as automatic accessibility auditing and indoor scanning and reconstruction using LiDAR.

\subsection{Home Accessibility/Safety Auditing Methods}
To improve the accessibility and safety of home spaces and improve the person-environment 'fit' \cite{law1996person}, a thorough assessment for potential risks based on residents' needs is required. This assessment---called \textit{home accessibility and safety auditing}---traditionally involved professional occupational therapists (OTs) with experience and insights about the challenges and remedies for home accessibility issues \cite{iwarsson_housing_1999}. To help OTs conduct standardized evaluations, checklists like \textit{WeHSA }\cite{clemson_home_1997}, \textit{Housing Enabler} \cite{iwarsson_housing_1999}, \textit{SAFER-Home} \cite{chiu2006factor}, and \textit{HEAVI} \cite{swenor_evaluation_2016} were developed and deployed. These checklists contain potentially hundreds of potential risks for OT's to monitor during home visits (\textit{e.g.,} uneven steps, bed too high, slippery floors).

However, healthcare system barriers like limited funding and lengthy insurance approval processes can impede on-site OT interventions \cite{renda_feasibility_2018}. In this case, occupant-oriented checklists, like the well-known HSSAT \cite{tomita_psychometrics_2014, horowitz_use_2016}, help people audit home spaces by themselves. Compared to the checklists for OTs, these consumer checklists \cite{fange_physical_1999,gray_subjective_2008,horvath_clinical_2013,gitlin_evaluating_2002} are designed to be more community/demographic specific (\textit{e.g.,} for older adults only), pictorial, and subjective. They usually contain detailed descriptions of potential risks and ways to overcome them. 

Besides checklists, which aim to discover problems in existing spaces, legal regulations also address home space accessibility. For example, the \textit{Fair Housing Act (FHA) Design Manual} \cite{housing_fair_1996} and \textit{the Americans with Disabilities Act (ADA) Design Guidelines}\cite{noauthor_ada_2010} present detailed design guidelines for home design and construction. Unlike home accessibility checklists, these guidelines are more specific about measurements of dimension and positioning of housing components, but they are less specific about specific communities/demographics and usually not as close a fit for personal needs.

In general, current home accessibility assessment practices require either manual measurements and ongoing checking or the participation of professional OTs. Such practices continue to pose barriers to home assessments due to residents' abilities, financial budget, and motivational factors. We address this gap by developing and evaluating RASSAR to enable reliable, fast, and always-available home auditing using smartphones, reducing the effort and potential resources needed to accomplish this vital task.

\subsection{Automatic Evaluation of Real-world Safety/Accessibility}
There has been a growing interest in using the latest sensing and computing technologies, such as crowdsourcing, indoor reconstruction, and computer vision, to improve safety and accessibility of the built environment. However, we observe an imbalanced focus on outdoor \textit{vs.} indoor spaces. One outdoor example is Project Sidewalk \cite{saha_project_2019,weld_deep_2019,sharif_experimental_2021}, which crowdsources annotations on street view data and uses deep learning models (such as ResNet) to detect target objects like curb ramps and inaccessible sidewalk conditions. Other works apply similar pipelines to pedestrian facilities \cite{luttrell_data_2022,lange_strategically_2021}and street bikeability \cite{ito_assessing_2021}. Most relevant to our work is Ayala-Alfaro \textit{et al.}'s research on indoor obstacle identification \cite{ayala-alfaro_automatic_2021}. While similar, our work provides user participation and verification, a wider range of accessibility issues, as well as customizability for people with different accessibility needs. 


In addition to image-based methods, researchers are using agent-based modeling \cite{fu_human-centric_2020}, graph-based methods \cite{dao_three-dimensional_2018,hashemi_indoor_2016} and point cloud data \cite{anjanappa_deep_2022,ayala-alfaro_automatic_2021,balado_automatic_2017,serna_urban_2013} to evaluate accessibility in built environments. For example, Fu \textit{et al.} \cite{fu_human-centric_2020} place virtual human agents into given 3D indoor scenes to interact with indoor objects in order to evaluate functional accessibility. Balado \textit{et al.} \cite{balado_automatic_2017} uses a MLS (Mobile Laser Scanner) to scan the facade of buildings and segment the point cloud data to detect potential accessibility issues at building entrances. Compared to image-based methods, these works are more difficult to apply at large scale due to the cost of specialized hardware and the complexity of data collection and analysis. 

In general, existing work requires massive amounts of data collection or specialized hardware to generate identified accessibility issues. In contrast, RASSAR empowers individuals to identify home accessibility issues tailored to their unique requirements using their smartphones. Our distinctive approach actively engages users in the AR-based scanning and evaluation processes.


\subsection{LiDAR-based Indoor Scanning \& Reconstruction}

In recent years, many mobile device manufacturers such as Apple, Samsung, and Huawei, have incorporated LiDAR sensors into smartphones. Apple’s iPhone 12 Pro, for example, was released in 2020 \cite{apple_apple_2020} and provides both hardware and software support for users to conduct scans and reconstruct indoor spaces. Although smartphone-based LiDAR capabilities cannot match professional devices, researcher evaluations demonstrate ample precision \cite{diaz_vilarino_3d_2022,luetzenburg_evaluation_2021,vogt_comparison_2021}. In May 2022, Apple released another indoor reconstruction API called RoomPlan \cite{apple_roomplan_2022}, which uses the camera, LiDAR sensor, and deep learning models to create real-time parameterized indoor models. This API simplifies the indoor reconstruction process and expands potential application scenarios by generating 3D models with dimension, position, and category. 


Although 3D reconstruction technology is rapidly advancing, two important gaps remain: the lack of focus on accessibility and safety in indoor reconstruction, and the inability to detect and locate accessibility-related objects in indoor spaces. RASSAR addresses these gaps by combining the mobile-based indoor scanning and reconstruction pipeline with accessibility rubrics and offering a custom, accessibility-focused object detection model.

\section{Design Process}
\label{sec:designProcess}
To design RASSAR, we conducted a three-stage iterative design process. We first built an \textit{initial technical prototype }to demonstrate feasibility. Then, we used our initial prototype as a design probe to conduct a \textit{formative user study} with five key stakeholder groups (Study 1). Finally, based on formative study findings, we built the \textit{current RASSAR system} with improved detection performance, rubric formation, user interaction, and interface design---and then performed both a technical evaluation (Study 2) and a user study (Study 3).

\subsection{Technical Prototype}
Informed by prior work in home accessibility assessment \cite{balado_automatic_2017,fu_human-centric_2020} and indoor reconstruction techniques \cite{9201064,apple_roomplan_2022}, we built a rapid technical prototype on an iPhone 13 Pro Max using \textit{Apple's RoomPlan API}~\cite{apple_roomplan_2022}. The prototype included three primary features: (1) a reconstruction of indoor spaces with accessibility-related items, (2) the detection of accessibility and safety issues in indoor scenes, and (3) an AR-based  visualization of and interaction with detected accessibility and safety issues.  We used this prototype to conduct controlled experiments of a single apartment to demonstrate technical feasibility and determine ideal scan conditions, such as varying levels of room tidiness, lighting, and moving speeds \cite{su_towards_2022}. 

Our initial findings suggest that scanning must be conducted at moderate speed (moving at ~0.5 meters/sec) while keeping the room tidy and well-lit. Under such conditions, the technical prototype's detection recall reached 90\%. The experiment also revealed deficiencies in object detection performance and UI limitations, which we later improved in the final RASSAR system (Section~\ref{sec:system}).

\subsection{\rev{Study 1:} Formative Study}
To examine the potential of semi-automatic accessibility and safety scanning with mobile phones and to solicit feedback of our initial technical prototype, we conducted a three-part formative user study with 18 participants drawn from five communities: wheelchair users (\textit{N=}8), families with young children (\textit{N=}3), people who are blind or low vision (\textit{N=}4), older adults (\textit{N=}6, including caregivers), and occupational therapists (\textit{N=}3); see \autoref{tab:demographics}.



\subsubsection{Participants}
Participants were recruited via email, advertisements to accessibility organizations, and social media as well as through snowball sampling. Before participating, all individuals filled out a screener on demographics and relationship(s) to our target communities. To ensure a diverse sample, we recruited at least three individuals from each community who were either self-identified members or caregivers. 

\begin{table}[htb!]
\centering
\label{table:demographics}
\caption{Our formative study included 18 participants. \textit{PID} stands for Participant Index. \textit{Older Adults} refers to those aged 65 and over. \textit{Children} to those families with children between 0 and 3 years old. \textit{BLV} to people who are blind or low vision. \textit{OT}  to occupational therapists. \textit{CG} to caregiver. As can be observed, identities/roles are not mutually exclusive.}
 \renewcommand{\arraystretch}{0.8}
 \resizebox{0.40\textwidth}{!}{
  \begin{tabular}{lcllll}
\hline
\textbf{PID} & \textbf{\begin{tabular}[c]{@{}c@{}}Wheelchair \\ User\end{tabular}} & \textbf{\begin{tabular}[c]{@{}l@{}}Older \\ Adults\end{tabular}} & \textbf{Children} & \textbf{BLV} & \textbf{OT}  \\ \hline
P1           & $\checkmark$                                                        &                                                                  &                   &              &              \\
\rowcolor[HTML]{EFEFEF} 
P2           & $\checkmark$                                                        &                                                                  &                   &              &              \\
P3           & $\checkmark$                                                        &                                                                  &                   &              &              \\
\rowcolor[HTML]{EFEFEF} 
P4           & $\checkmark$                                                        &                                                                  &                   &              &              \\
P5           & $\checkmark$                                                        &                                                                  &                   &              &              \\
\rowcolor[HTML]{EFEFEF} 
P6           & $\checkmark$                                                        &                                                                  &                   &              &              \\
P7           &                                                                     &                                                                  & CG                &              &              \\
\rowcolor[HTML]{EFEFEF} 
P8           & $\checkmark$                                                        & $\checkmark$                                                     &                   &              &              \\
P9           &                                                                     & CG                                                               & CG                &              &              \\
\rowcolor[HTML]{EFEFEF} 
P10          & $\checkmark$                                                        & $\checkmark$                                                     &                   &              &              \\
P11          &                                                                     &                                                                  &                   &              & $\checkmark$ \\
\rowcolor[HTML]{EFEFEF} 
P12          &                                                                     &                                                                  &                   &              & $\checkmark$ \\
P13          &                                                                     & $\checkmark$                                                     &                   & $\checkmark$ &              \\
\rowcolor[HTML]{EFEFEF} 
P14          &                                                                     & $\checkmark$                                                     &                   & $\checkmark$ &              \\
P15          &                                                                     & $\checkmark$                                                     &                   & $\checkmark$ &              \\
\rowcolor[HTML]{EFEFEF} 
P16          &                                                                     &                                                                  & CG                &              &              \\
P17          &                                                                     &                                                                  &                   &              & $\checkmark$ \\
\rowcolor[HTML]{EFEFEF} 
P18          &                                                                     &                                                                  &                   & $\checkmark$ &              \\ \hline
\end{tabular}}
\end{table}

\subsubsection{Procedure}
We conducted a three-part, qualitative study. Part 1 addressed current practices, and Parts 2 and 3 focused on our RASSAR design probe. Specifically, in Part 1, we asked participants about their indoor accessibility and safety needs, their current practices for assessing those needs, and challenges therein. For Part 2, we showed participants videos of RASSAR scanning an apartment and solicited suggestions, concerns, and expected usage scenarios in their own lives. Finally, in Part 3, we showed RASSAR UI mockups \rev{as design probes} and asked participants \rev{for preferences}: how to encode new accessibility/safety issues into RASSAR (\textit{e.g.,} rubric customization), how to guide the user-conducted scan, and how to provide feedback for scan results and errors. Sessions lasted between 40-70 minutes (\textit{Avg=}52 mins), and all but one was conducted over Zoom. The lead author conducted all 18 sessions. For reference, we include the full interview protocol and design probe in the supplementary files.






\subsubsection{Data and analysis}\label{subsection:formativeresults}
We audio and video transcribed all sessions via Rev~\cite{revcom_transcribe_nodate}. For analysis, we conducted reflexive thematic coding \cite{braun_reflecting_2019}. The first author, who also conducted the interviews, reviewed all transcripts to develop an initial codebook. The first four authors then discussed the initial codebook, iteratively resolved disagreements and developed a full version of the codebook collaboratively. The first author then applied the finalized codebook to all transcripts. After coding ended, the first four authors met and discussed general themes and findings.

\subsubsection{Findings} We highlight four key findings below related to indoor accessibility and safety needs, current practices and challenges in assessing indoor accessibility and safety, reactions to \textit{RASSAR}, and our design probe results.

\textbf{Overall reactions to RASSAR.} Most participants (\textit{N=}16) held favorable opinions about the  RASSAR prototype due to its measurement and documentation features, its ability to help prepare a home for visitors with accessibility needs, and its ability to customize accessibility issues. The two participants who held unfavorable or neutral opinions stated, ``\textit{I don't see the point since I can do these screenings by myself}'' and ``\textit{I don't know.}''

Participants described a variety of usage scenarios for RASSAR. Six participants mentioned how RASSAR could help people gain knowledge of physical spaces before they actually visit them. For example, P18 (BLV) said, \textit{“I think it could really help people have more confidence when they go into a room that they're not familiar with or a new space.”} Another commonly raised use-case was facilitating renovation or real-estate viewings; for example, P4 (wheelchair user) said, ``\textit{I could even see this as being incredibly helpful to send to contractors and architects and designers.}'' Finally, participants were excited about ways RASSAR could facilitate accommodation during travel, \rev{since hotel or Airbnb managers can better communicate accessibility situations with RASSAR scan results}. 

\textbf{People’s unique indoor accessibility needs.}
Our participants flagged common sources of indoor safety and accessibility issues, such as stairs (\textit{N=}14), doors (\textit{N=}9), and floors (\textit{N=}9), which also commonly appear in accessibility checklists \cite{tomita_psychometrics_2014,horowitz_use_2016, rebuilding_together_safe_nodate}. However, we also found key differences in concerns across communities. For example, no wheelchair user mentioned concern about hazardous items like sharp furniture corners or knives, while all other communities (BLV \textit{N=}3, older adults \textit{N=}4, children \textit{N=}2, OT \textit{N=}1) stressed it in interviews. This highlights the need for  RASSAR to be customizable to address different abilities and needs.

\textbf{Current practices and challenges.} None of our participants with accessibility needs previously chose to use accessibility checklists when auditing indoor spaces. They instead relied on their own experiences and formed methods through practice. The most commonly used methods were exploring the space (\textit{N=}6), creating mind maps of potential issues (\textit{N=}5), and asking questions about the space (\textit{N=}4). Caregivers typically put themselves in care receivers' shoes (\textit{N=}2) to better identify potential issues. Current practices also posed inherent challenges, including other people's limited understanding of accessibility (\textit{N=}3) and social awkwardness (\textit{N=}3). 

For example, P10 said, ``\textit{People mean well, they really do. [But] until you actually have to live it or be exposed to it on a routine basis, you just really don't understand [our accessibility challenges].}'' \rev{Similarly, P6 said, ``\textit{The main challenge is explaining to people how to look at it from my perspective.}'' In this case, audits should be conducted on-site by individuals with accessibility needs themselves for reliability, requiring extra time or money to resolve or circumvent issues. P8 recalled an instance of calling a restaurant to confirm accessibility, only to find it inaccessible upon arrival, forcing her to choose a nearby alternative.} We maintain that these challenges can be mitigated by indoor auditing methods such as RASSAR, \rev{which provides a holistic digital scan of space that can be evaluated remotely based on general or personal requirements.}

\begin{figure}[h]
  \centering
  \includegraphics[width=\linewidth]{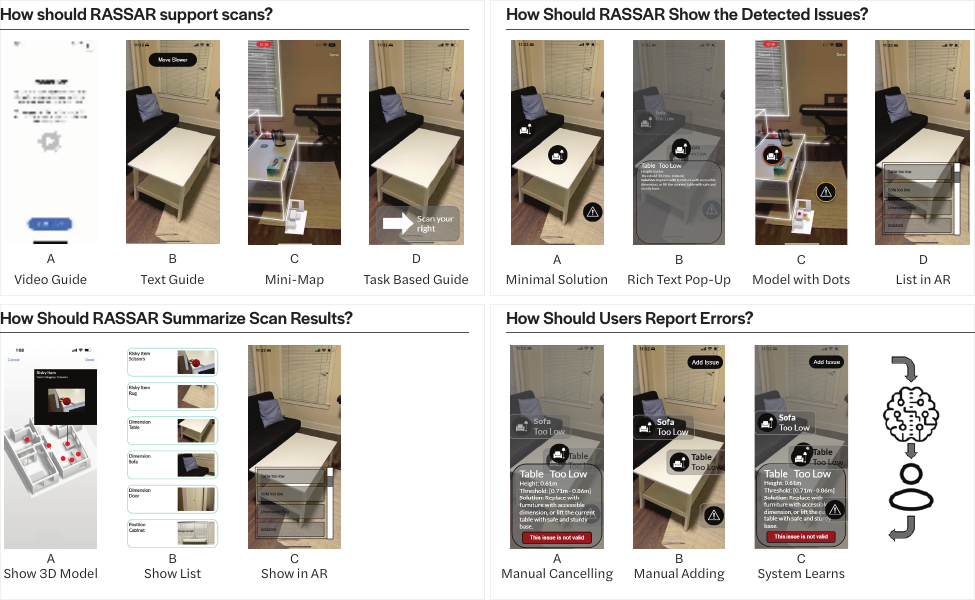}
  \caption{Design probes for RASSAR's interaction and interface design. For each question, we provide three or four options as interface mock-ups.}
  \Description[Four questions and their respective options as interface mock-ups]{Four cluster of diagrams showing the design probes. Each contains a question and 3 - 4 options indicated with a UI mock-up and a name.}
  \label{fig:probes}
\end{figure}

\textbf{Design improvements for RASSAR rubrics and UI.} 
\label{para/design improvements}We presented participants with a list of accessibility and safety issues derived from literature to solicit feedback. Most participants were satisfied, while some suggested additions. Based on feedback, we modified and finalized RASSAR's auditing rubrics (\textit{e.g.,} added new issues like bed height). We also provided UI mockups to solicit participants' preferences about RASSAR's scan experience (\autoref{fig:probes}). Participants preferred having text hints (\textit{N=}15) and mini-maps (\textit{N=}11) for scanning support, an interactive rich text pop-up layer (\textit{N=}13) for detected issue visualization, \rev{an interactive 3D model} (\textit{N=}9), and a list of issues to review (\textit{N=}10) in a post-scan summary. \rev{As for methods of error reporting, most participants (\textit{N=}15) preferred that the system learn their needs over time.}

\section{The RASSAR System}
\label{sec:system}
\begin{figure*}[h]
  \centering
  \includegraphics[width=\textwidth]{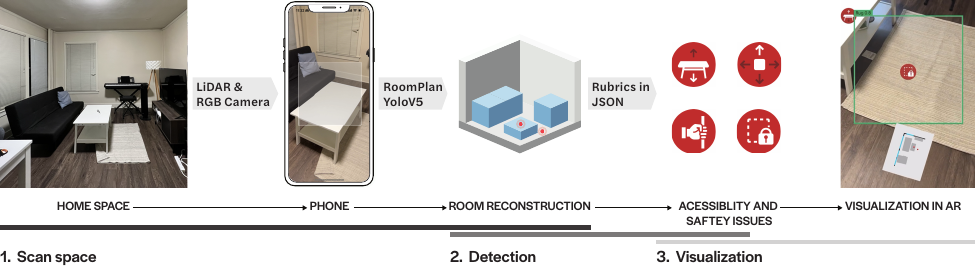}
  \caption{RASSAR system overview. RASSAR (1) scans and reconstructs a  home space, (2) detects accessibility and safety issues in the space, and (3) visualizes real-time results in AR.}
  \label{fig:systemOverview}
  \Description[The overview diagram of \textit{RASSAR}'s technical pipeline.]{The system overview of RASSAR. On the top there are five images linked with text in arrow. From left to right these images are: a picture of a living room, a phone with RASSAR’s scanning interface on it, a simplified model view of indoor reconstruction, four icons used to show detected accessibility issues in RASSAR, RASSAR’s actual UI visualization of detected issue and mini map.}
\end{figure*}

Informed by our rapid prototype and formative study findings, we created a revised RASSAR system (\autoref{fig:systemOverview}), which uses LiDAR and real-time CV to perform a parametric reconstruction of indoor spaces, automatically identify safety and accessibility problem using a custom JSON rubric, and visualize issues both in AR and via an interactive 3D summary view. 

The new prototype includes four key enhancements: (1) \textit{Improved feedback:} all detected accessibility and safety issues are visualized in real time with multiple UIs, including a pop-up icon in the AR interface, descriptive layers, and post-hoc summarization in the 3D model. This empowers users to actively engage with the system to explore, refine, and summarize scan outcomes. (2) \textit{Reduced scanning effort:} RASSAR users can learn to scan rooms with both textual and mini-map hints, and the app is designed to help users scan spaces without having to move close to potential objects of interest, thereby reducing scanning effort. (3) \textit{Customizability:} RASSAR users can tailor the scanning process by selecting different accessibility communities to filter detection rubrics. They can also manually remove detected issues that fail to match their specific needs. (4) \textit{BLV support:} \rev{To assist BLV users, each of the primary user interactions—selecting target communities, scanning, object identification, and summary of results—are supported by real-time audio feedback. All user interface components, such as buttons and text labels, are compatible with VoiceOver.} Below, we describe RASSAR's scanning, detection, and visualization process. 


\subsection{Scan and Reconstruct Home Space}
In Step 1 of RASSAR's technical pipeline, we reconstruct a scanned home space into a parametric 3D model that includes object category, dimension, and position information. To create a reconstruction with access/safety information about both \textit{macro} objects (\textit{e.g.,} furniture) and \textit{micro} objects (\textit{e.g.,} electric sockets), we combine \textit{Apple's RoomPlan API} \cite{apple_roomplan_2022} and a customized \textit{YOLOV5} model \cite{jocher_ultralyticsyolov5_2022}.

\subsubsection{Parametric eeconstruction of major indoor components}
The RoomPlan API is the backbone of our indoor reconstruction process. RoomPlan, which relies on both RGB camera and LiDAR sensor data, provides real-time spatial and dimension information for major indoor components. Currently, RoomPlan detects the following object categories: \textit{bathtub}, \textit{door}, \textit{opening}, \textit{wall}, \textit{window}, \textit{bed}, \textit{chair}, \textit{sink}, \textit{sofa}, \textit{stairs}, \textit{storage}, \textit{table}, \textit{television}, and \textit{toilet}. However, this reconstruction is incomplete; it includes only walls, doors and furniture but neglects smaller objects that could also pose accessibility and safety challenges.

\subsubsection{Object detection for smaller indoor objects}
\label{subsubsec:object detection}
Based on our literature review, we identified many issues related to smaller objects, such as \textit{grab bars}, \textit{electric sockets}, and \textit{sharp objects}. Since no existing pre-trained model is tailored for accessibility-related indoor objects, we trained our own object detection model to complement our room reconstruction pipeline. \rev{Informed by \cite{noauthor_ada_2010,housing_fair_1996, tomita_psychometrics_2014,horowitz_use_2016, KidsClinicSafety,nguyen2021reducing}, we selected nine initial categories of smaller indoor items that are common problems and }collected a customized dataset with 2533 images and 3943 annotations (\autoref{tab:dataset}). About one third of these images come from the \textit{Open Images} dataset \cite{OpenImages}, while the remainder comes from \textit{Microsoft's Bing} \cite{Microsoft}. The first author manually annotated annotated all instances of the nine object categories (\autoref{tab:dataset}) with bounding boxes.

Using the dataset, we trained a state-of-the-art computer vision model, YOLOV5\cite{jocher_ultralyticsyolov5_2022}. We specifically selected \textit{YOLOV5-m} (41 MB) due to its fast detection speed (224ms inference time with CPU \cite{yolov5github}) and good performance on the baseline COCO dataset (0.63 for mAP@0.5\footnote{mAP@0.5 represents the \textit{Mean Average Precision} at an \textit{Intersection over Union} (IoU) threshold of 0.5}). In offline experiments, we found that a larger model would slightly increase performance (0.66 for mAP@0.5) but almost double detection speed (430ms). We trained our model on an NVIDIA GTX 3080 and Ubuntu 20.04.2 LTS using 900 epochs. After obtaining the best weights, we converted them into the CoreML format (.mlmodel) \cite{apple_core_nodate} by adding a non-maximum suppression layer \cite{jocher_ultralyticsyolov5_2022}. We randomly sampled our dataset into training, validation, and test sets containing 70\%, 15\%, and 15\% of the data, respectively. The model performance is shown in Table~\ref{tab:training}. To facilitate open science, we have open sourced the annotated dataset \footnote{\url{https://github.com/makeabilitylab/RASSAR}}. 

\begin{table}[htb!]
\caption{The number of images and annotations for training RASSAR’s customized YOLO model. Some images contained multiple objects, which is why the annotation count exceeds the image count.}
  \label{tab:dataset}
  \renewcommand{\arraystretch}{1.05}
    \begin{tabular}{lll}
\hline
\textbf{Object}            & \textbf{\begin{tabular}[c]{@{}l@{}}Images\end{tabular}} & \textbf{\begin{tabular}[c]{@{}l@{}}Annotations\end{tabular}} \\ \hline
Door handle                & 370                                                            & 530                                                                 \\
Electric socket            & 181                                                            & 370                                                                 \\
Grab Bar                   & 395                                                            & 503                                                                 \\
Knife                      & 451                                                            & 622                                                                 \\
Medication                 & 325                                                            & 688                                                                 \\
{\color[HTML]{000000} Rug} & {\color[HTML]{000000} 377}                                     & {\color[HTML]{000000} 470}                                          \\
Scissors                   & 226                                                            & 270                                                                 \\
Smoke alarm                & 176                                                            & 191                                                                 \\
Light switch               & 138                                                            & 299                                                                 \\ \hline
\textbf{Total}             & \textbf{2533}                                                  & \textbf{3943}                                                       \\ \hline
\end{tabular}
\end{table}
\begin{table}[htb!]
\centering
\caption{Performance of our trained model for each object category. mAP is mean Average Precision, and 0.5 is a common threshold to determine the effective intersection over union (IoU).}
  \label{tab:training}
\renewcommand{\arraystretch}{1.05}
\resizebox{0.45\textwidth}{!}{
\begin{tabular}{lllll}
\hline
\textbf{Class}  & \textbf{Target} & \textbf{Precision} & \textbf{Recall} & \textbf{mAP@0.5} \\ \hline
All             & 587             & 0.744              & 0.865           & 0.869            \\
\rowcolor[HTML]{EFEFEF} 
Door Handle     & 66              & 0.601              & 0.803           & 0.746            \\
Electric Socket & 30              & 0.719              & 0.867           & 0.877            \\
\rowcolor[HTML]{EFEFEF} 
Grab Bar        & 104             & 0.777              & 0.952           & 0.972            \\
Knife           & 93              & 0.682              & 0.71            & 0.756            \\
\rowcolor[HTML]{EFEFEF} 
Medication      & 113             & 0.831              & 0.938           & 0.944            \\
Rug             & 60              & 0.965              & 0.983           & 0.994            \\
\rowcolor[HTML]{EFEFEF} 
Scissors        & 60              & 0.775              & 0.85            & 0.887            \\
Smoke Alarm     & 29              & 0.765              & 0.931           & 0.873            \\
\rowcolor[HTML]{EFEFEF} 
Switch          & 32              & 0.579              & 0.75            & 0.767            \\ \hline
\end{tabular}}
\end{table}

During the scanning process, the custom YOLOV5 model runs continuously on the camera feed to detect the nine accessibility-related objects in the scanned space. Since YOLO detection results are 2D bounding boxes on images instead of 3D coordinates in a room, we perform raycasting \cite{apple_raycast_2023} to convert the 2D location from the center of the YOLO bounding box into 3D coordinates in physical space. To improve accuracy and reduce noise and outliers, we smooth out the raycasting results by averaging across multiple frames and setting up filtering thresholds. For a YOLO-detected object to be considered valid, it needs five raycasting results with location offset of fewer than 0.3 meters and YOLO detection confidence scores greater than 0.65.

Combining the results of RoomPlan and YOLOV5, RASSAR produces a real-time indoor reconstruction that includes category, dimension and position information of both larger barriers and many smaller indoor objects. This room reconstruction provides a solid basis for accessibility and safety auditing.

\subsection{Detection of Accessibility and Safety Issues}
In Step 2 of RASSAR's pipeline, we filter and identify relevant accessibility and safety issues based on the selected stakeholder group(s) and a customizable rubric.

\subsubsection{Rubric formation}
\label{subsubsec:rubrics formation}
\begin{figure*}[h]
  \centering
  \includegraphics[width=\textwidth]{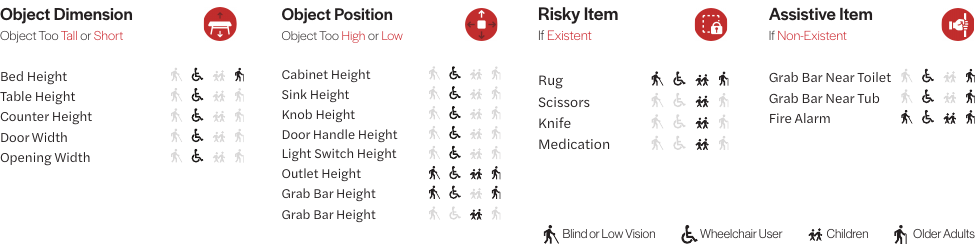}
  \caption{Informed by literature and our formative study, RASSAR can detect 20 types of accessibility and safety issues across four categories:\textit{ object dimension}, \textit{object position}, \textit{risky item}, and \textit{assistive item}. Each issue has relevance to specific accessibility communities, marked with black icons. We acknowledge that safety/accessibility issues can be fluid marked not just by (dis)ability but fatigue, time-of-day, \textit{etc.} and that individuals may not map exactly to these categories. Our custom JSON-based rubric could allow for precise individual specification in the future (\textit{e.g.,} with a custom authoring interface.}
  \label{fig:rubrics}
  \Description[RASSAR's 20 types of accessibility and safety issues, falling into four categories.]{All 20 accessibility issues classified by their categories. Each issue show four icons indicating their respective relevance to accessibility communities.}
\end{figure*}

RASSAR's auditing rubric is drawn from \textit{ADA Design Guidelines} \cite{noauthor_ada_2010}, the \textit{Home Safety Self Assessment Tool} (HSSAT) \cite{tomita_psychometrics_2014,horowitz_use_2016}, the \textit{US Fair Housing Act Design Manual} \cite{housing_fair_1996,steinfeld_home_1998}, and other sources \cite{taira_aging_2014, osborne_home_2008}. Creating the default rubrics for each stakeholder group was iterative and additionally informed by Study 1. The final rubrics include 20 issues across four categories (\autoref{fig:teaser}.2, \autoref{fig:rubrics}): \textit{object dimension} (\textit{e.g.,} high table height), \textit{object position} (\textit{e.g.,} out-of-reach light switch), \textit{risky items} (\textit{e.g.,} presence of a sharp object like scissors), and \textit{lack of assistive items} (\textit{e.g.,} missing bathtub grab bars). For more details, see \Cref{appendix:rubrics}. When RASSAR starts, users can select one or more target communities, which will load the relevant rubric(s).


\subsubsection{JSON-encoded rubrics}
\label{json formatting}
To facilitate both automated screening of room reconstruction and individual customization, we transformed our original text-based rubrics into \textit{JavaScript Object Notation} (JSON) format. Each JSON-formatted rubric encompasses essential details, such as the target object category (\textit{e.g.}, table), its relevant user community (\textit{e.g.}, wheelchair users), \rev{the dependent other object when rubric involve multiple objects (\textit{e.g.}, tub for the issue of \textit{No Grab Bar Near Tub})}, and the violation criteria (\textit{e.g.}, dimensions less than 68 cm, and relative distance more than 70 cm away). Moreover, we enriched these JSON files with supplementary information, including warning messages, issue descriptions, suggestions, and information sources. This additional context aids users in comprehending detected issues and assists in the removal of invalid or irrelevant results. See \Cref{appendix:json} for an example. 



\subsection{Visualization of scans}
Finally, in Step 3, RASSAR provides a real-time 3D reconstruction to aid scanning, a visualization of identified issues in AR, and a post-hoc summary of scan results.


\begin{figure}[h]
  \centering
  \includegraphics[width=1\linewidth]{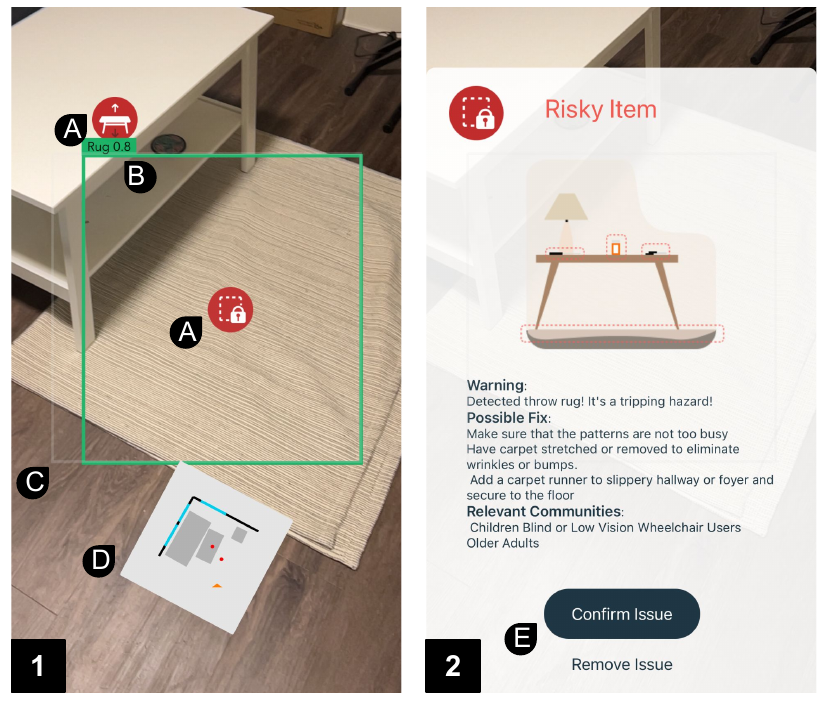}
  \caption{RASSAR's AR-based scanning interface and a detailed view of a detected issue. (1) During a scan, RASSAR shows detected problems in real-time via AR overlays, including (a) red spheres, which can be selected to view more information and to confirm/delete detection and (b) CV-based detections with green bounding boxes, a text label, and confidence score. To aid understanding of the CV field-of-view, we draw a (c) gray bounding box. We also show a mini-map (d) that adapts to users' orientation/position with real-time reconstruction results. (2) The user can click on identified issues (a) to view more information, see recommended solutions, and to (e) confirm/delete problems.}
  \Description[Two UI screenshots side by side showing RASSAR's visualization of scan process.]{Two UI screenshots. The left one shows RASSAR’s scanning interface with red pop-up icons indicating detected issues, green bounding box showing a rug detected by YOLO, and a mini-map showing the room reconstruction. The right one shows details of one detected issue, which is the throw rug on floor. It also include information like why this issue is risky and what’s the possible fixing solutions.}
\label{fig:interface}
\end{figure}

\subsubsection{Facilitating user scanning}

The RASSAR interface employs several user feedback methods to facilitate scanning. First, real-time room reconstruction progress is visually represented through a dynamic mini-map (\autoref{fig:interface}d). This mini-map adapts to the user's orientation and shows real-time reconstruction progress with distinct visual cues (\textit{e.g.,} black lines signify walls, yellow lines denote doors, and the orange triangle indicates the user's current position and direction). Second, to enhance the capture of smaller indoor objects from a distance, RASSAR optimizes its camera feed by utilizing only the central part of the screen as input to the YOLO model. Users are guided by a subtle, dimmed white box (\autoref{fig:interface}b). Third, all object detection results are presented in YOLO-style bounding boxes (\autoref{fig:interface}c). Finally, users can also enable audio feedback during the scanning process, which verbally guides the scan with instructions such as ``\textit{Please point camera at top and bottom of wall to initialize}'', ``\textit{Please slow down}'', and ``\textit{Please step away from wall.}'' All major indoor objects, such as doors, windows, tables, and sofas, are read out when detected.
\subsubsection{AR visualization of detected issues}
\label{subsubsec:visualization in AR}
We also overlay detections in real-time using AR. The four classes of accessibility issues (\textit{object dimension}, \textit{object position}, \textit{risky item}, and \textit{lack of assistive item}, see \autoref{fig:rubrics}) are encoded and shown via four corresponding pop-up icons. Icons are clickable to inspect detailed information (\autoref{fig:interface}.2) and verify if the issue is valid. If not, the user can remove the issue (\autoref{fig:interface}e). With audio support enabled, RASSAR also verbalizes when an accessibility or safety issue is detected, such as ``\textit{Too narrow door opening is detected!}'', ``\textit{The bed is too high!}'', and ``\textit{No grab bar detected near toilet!}''


\subsubsection{Post-scan interactive 3D summary}
After a scan, \rev{RASSAR provides a brief summary of detected issues, including issue counts and their category---which can be read out with VoiceOver} RASSAR \rev{then} shows an interactive 3D model of results where indoor components and objects are represented as geometries and detected issues are shown in red. The user can pan, zoom and move the 3D model to inspect more closely or tap on any object to see more details. The lower half of the UI shows details about user-selected objects, including object category and dimension. The user can also delete detected issues or export all scan results into a JSON file.

\begin{figure}[h]
    
  \centering
  \includegraphics[width=1\linewidth]{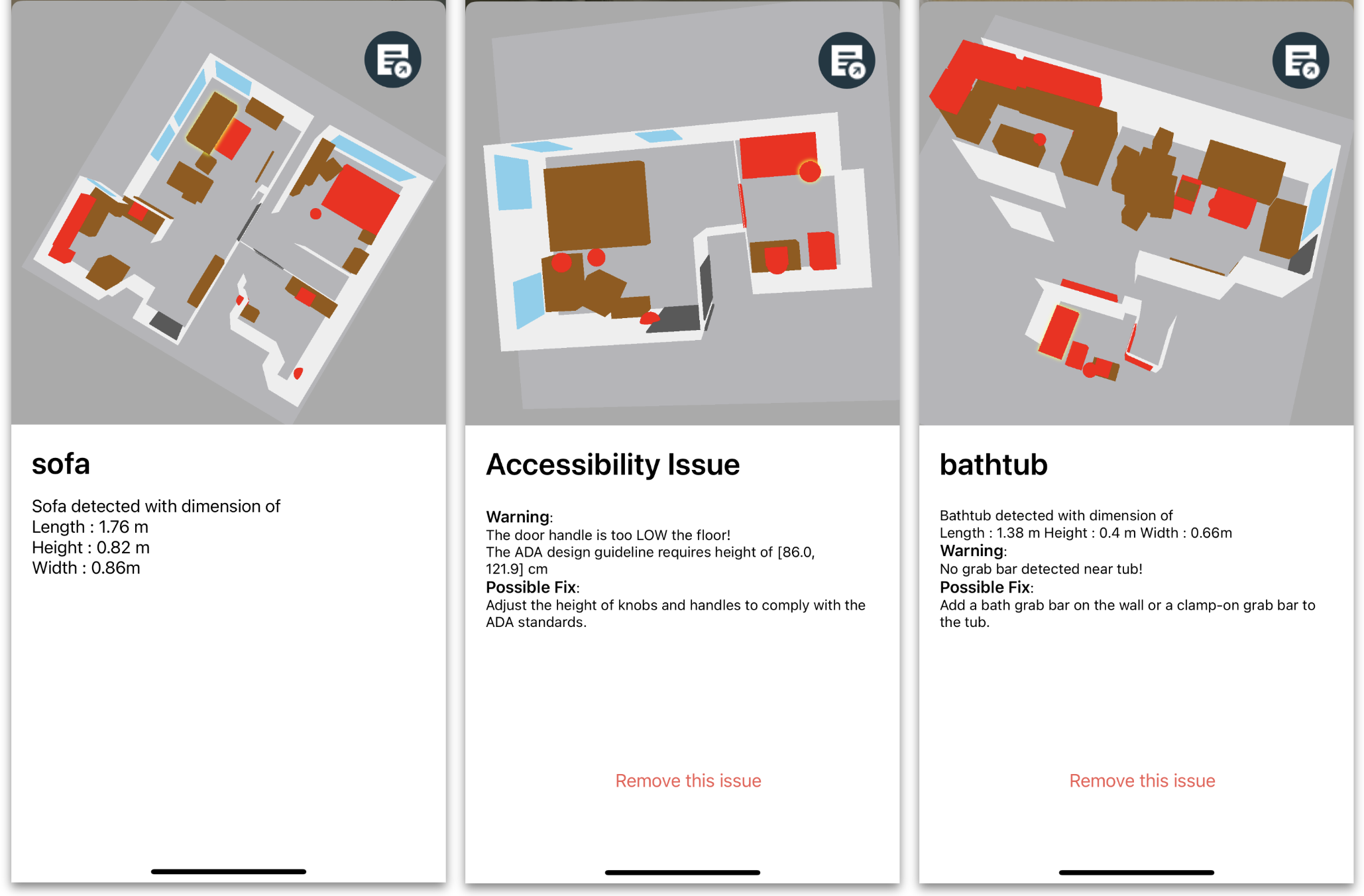}
  \caption{Three examples of the post-scan interactive summary of results. A user can interact with the 3D reconstruction, inspect detailed information about objects or issues, and remove any errant or disagreed upon issues. The top-right button lets user export scan results to a JSON file.}
  \Description[Three of RASSAR's summary UI example.]{Three of RASSAR's summary UI example. Each include a 3D model view and a text box including elaboration of certain selected object. The first UI shows a top view of an apartment layout, with sofa selected. The second UI shows a top view of a bedroom and its restroom, with an accessibility issue about door handle being too low selected. The third UI shows a living room and a restroom, with an accessibility issue of bathtub lack grab bar selected.}
\label{fig:summary}
\end{figure}


\rev{
\section{Study 2: Technical Evaluation}\label{sec:study2}
}
Prior work in automatic accessibility auditing typically includes technical performance evaluation "case studies" in 3-4 real-world environments \cite{balado_automatic_2017,deng_semantic_2022,diaz-vilarino_obstacle-aware_2019}. Expanding on this approach, we performed technical evaluations of RASSAR in 10 home spaces, including seven apartments and three houses of varying sizes and layouts (\Cref{tab:evaluationresults}). \rev{The first author, an experienced accessibility researcher and expert RASSAR tool user, conducted all scans.}

\vspace{12pt}
\subsection{Experiment Procedure}
\label{study2procedure}
For each indoor space, we performed a three-step process. First, we conducted a \textit{manual audit} to identify any existing accessibility and safety issues using RASSAR's accessibility rubrics (\Cref{appendix:rubrics}). \rev{We inspected and measured each indoor space, used the rubric for assessment, and took additional notes about found issues\footnote{We excluded the \textit{`Knob Height'} issue from this study to avoid imbalance of issues in the manual inspection results.}.} Second, we \textit{scanned each space with RASSAR}, which we recorded and exported to JSON format. Finally, we \textit{compared the scan results} with the manual audit data. We repeated the last two steps three times, resulting in 30 RASSAR scans across the ten indoor spaces. 

\subsection{Evaluation Metrics}
\label{subsec:metrics}
To evaluate RASSAR, we used three primary evaluation measures: \textit{detection performance}, \textit{scanning consistency}, and \textit{scanning time}.

\subsubsection{Detection performance} For detection performance, we calculate the number of true positives (\textit{TP}), false positives (\textit{FP}), and false negatives (\textit{FN}) based on whether RASSAR successfully detected an issue listed in the RASSAR accessibility rubric, reported an issue that did not appear in the rubric, or failed to detect an issue. Notably, we did not calculate true negatives since it entail RASSAR not reporting on accessible and safe items, which is not meaningful for this evaluation. 

Based on TP, FP, and FN counts, we calculate four evaluation metrics, including \textbf{precision} (calculated as $\frac{TP}{TP+FP}$), 
\textbf{recall} (calculated as $\frac{TP}{TP+FN}$)
, \textbf{F1 score} \cite{powers2020evaluation}, and \textbf{accuracy} (calculated as $\frac{TP}{TP+FP+FN}$). These metrics assess the quality of RASSAR's output, how many issues RASSAR missed, and RASSAR's ability to capture existing issues and avoid false alarms. 

\subsubsection{Scanning consistency}

To examine RASSAR's consistency across scans (three per indoor space), we use Krippendorff's alpha \cite{krippendorff2011computing}, a common measure to assess agreement level among multiple raters. Specifically, we considered each scan a distinct rater and treated each space's accessibility and safety issue as a scoring task.

\subsubsection{Scanning time}
Finally, we computed the average scanning time for each indoor space using screen recordings of scans. 

\subsection{Evaluation Results}\label{subsec:evalresults}

\begin{figure}
  \centering
  \includegraphics[width=0.5\textwidth]{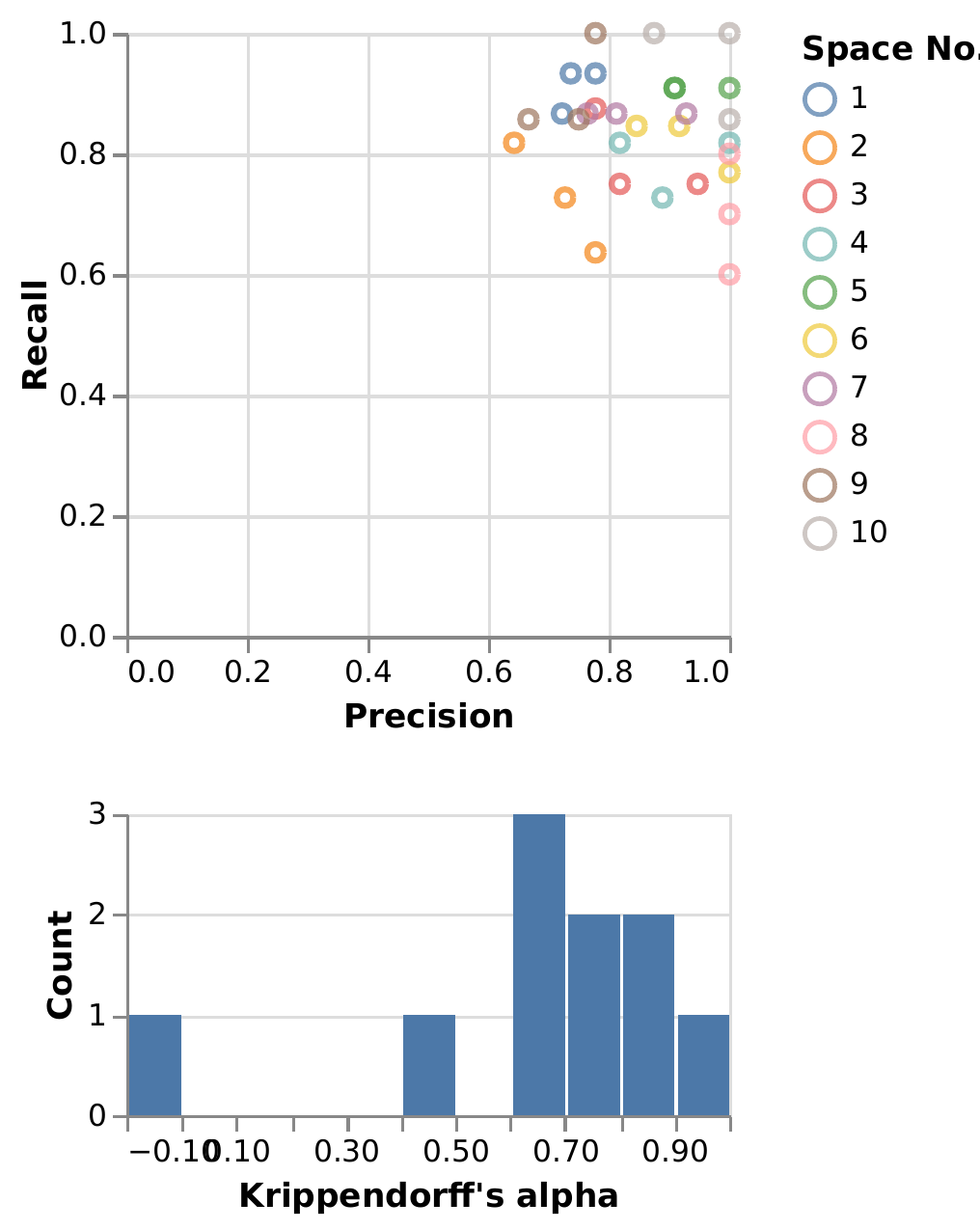}
  \caption{RASSAR's technical evaluation performance over ten indoor spaces. (Top) A scatter plot of precision and recall over 30 scans in 10 home spaces. (Bottom) A histogram of Krippendorff's alpha values across 10 home spaces (3 scans each).}
  \label{fig:distribution}
  \Description[Visualization of RASSAR’s technical evaluation results.]{Visualization of RASSAR’s technical evaluation results. Upper plot shows a scatterplot of precision and recall of all 30 scans. All dots are in the upper-right corner of the scatterplot, showing that all results have high precision and recall. Lower plot shows a histogram of Krippendorff’s Alpha values. Most results are in the range of 0.4 – 1.0.}
\end{figure}

\begin{table*}[]
\centering
    \caption{Information about the ten home spaces and RASSAR's performance results.Scanning time was measured in seconds. \rev{Precision, recall, accuracy, F1 score and Krippendorff's Alpha are described in \Cref{subsec:evalresults}. Performance metrics are averaged for each space. Raw scan performance data can be found in \Cref{appendix:eval}}}
     \label{tab:evaluationresults}
\begin{tabular}{llrr|rrrrrr}
\hline
\textbf{ID} & \textbf{\begin{tabular}[c]{@{}l@{}}Home \\ type\end{tabular}} & \multicolumn{1}{l}{\textbf{\begin{tabular}[c]{@{}l@{}}Size \\ (sqm)\end{tabular}}} & \multicolumn{1}{l|}{\textbf{\begin{tabular}[c]{@{}l@{}}Rooms \\ Scanned\end{tabular}}} & \multicolumn{1}{l}{\textbf{\begin{tabular}[c]{@{}l@{}}Average\\ Precision\end{tabular}}} & \multicolumn{1}{l}{\textbf{\begin{tabular}[c]{@{}l@{}}Average\\ Recall\end{tabular}}} & \multicolumn{1}{l}{\textbf{\begin{tabular}[c]{@{}l@{}}Average\\ F1 Score\end{tabular}}} & \multicolumn{1}{l}{\textbf{\begin{tabular}[c]{@{}l@{}}Average\\ Accuracy\end{tabular}}} & \multicolumn{1}{l}{\textbf{\begin{tabular}[c]{@{}l@{}}Krippendorf \\ Alpha\end{tabular}}} & \multicolumn{1}{l}{\textbf{\begin{tabular}[c]{@{}l@{}}Scan \\ Time\end{tabular}}} \\ \hline
S1          & Apartment                                                     & 65                                                                                 & 3                                                                                      & 0.75                                                                                     & 0.91                                                                                  & 0.82                                                                                    & 0.70                                                                                    & 0.73                                                                                      & 113                                                                               \\
\rowcolor[HTML]{EFEFEF} 
S2          & Apartment                                                     & 63                                                                                 & 2                                                                                      & 0.72                                                                                     & 0.73                                                                                  & 0.72                                                                                    & 0.56                                                                                    & 0.7                                                                                       & 120                                                                               \\
S3          & House                                                         & 45                                                                                 & 4                                                                                      & 0.85                                                                                     & 0.79                                                                                  & 0.81                                                                                    & 0.69                                                                                    & 0.67                                                                                      & 148                                                                               \\
\rowcolor[HTML]{EFEFEF} 
S4          & Apartment                                                     & 55                                                                                 & 3                                                                                      & 0.90                                                                                     & 0.79                                                                                  & 0.84                                                                                    & 0.73                                                                                    & 0.82                                                                                      & 80                                                                                \\
S5          & Apartment                                                     & 50                                                                                 & 3                                                                                      & 0.94                                                                                     & 0.91                                                                                  & 0.92                                                                                    & 0.86                                                                                    & 1                                                                                         & 84                                                                                \\
\rowcolor[HTML]{EFEFEF} 
S6          & Apartment                                                     & 90                                                                                 & 4                                                                                      & 0.92                                                                                     & 0.82                                                                                  & 0.87                                                                                    & 0.76                                                                                    & 0.83                                                                                      & 125                                                                               \\
S7          & Apartment                                                     & 65                                                                                 & 3                                                                                      & 0.84                                                                                     & 0.87                                                                                  & 0.85                                                                                    & 0.74                                                                                    & 0.62                                                                                      & 96                                                                                \\
\rowcolor[HTML]{EFEFEF} 
S8          & Apartment                                                     & 50                                                                                 & 3                                                                                      & 1.00                                                                                     & 0.70                                                                                  & 0.82                                                                                    & 0.70                                                                                    & 0.69                                                                                      & 80                                                                                \\
S9          & House                                                         & 24                                                                                 & 2                                                                                      & 0.73                                                                                     & 0.90                                                                                  & 0.81                                                                                    & 0.68                                                                                    & -0.05                                                                                     & 53                                                                                \\
\rowcolor[HTML]{EFEFEF} 
S10         & House                                                         & 60                                                                                 & 3                                                                                      & 0.92                                                                                     & 0.88                                                                                  & 0.90                                                                                    & 0.82                                                                                    & 0.43                                                                                      & 100                                                                               \\ \hline
\multicolumn{4}{r|}{\textbf{Average}}                                                                                                                                                                                                                     & \textbf{0.86}                                                                            & \textbf{0.83}                                                                         & \textbf{0.84}                                                                           & \textbf{0.72}                                                                           & \textbf{0.64}                                                                             & \textbf{99.9}                                                                     \\ \hline
\end{tabular}
\end{table*}
\Cref{tab:evaluationresults} and \Cref{fig:distribution} shows RASSAR evaluation results. \rev{For readability, we present averaged performance stats for each space. The full results can be found in the appendix: \Cref{tab:evaluationresultsFULL}.} Overall, our analysis shows that RASSAR can effectively identify indoor accessibility and safety issues, with an average precision of 0.86, recall of 0.83, F1 score of 0.84, and accuracy of 0.72. We observe that most Krippendorff's alpha values fall in the range between 0.4 and 0.7, which indicates "substantial consistency" \cite{hughes2021krippendorffsalpha}). Finally, RASSAR scanning took 99.9 seconds on average (\textit{SD=}27.4), much faster than manual auditing, which took the lead author approximately 10 mins/space. 

\subsection{Performance Analysis}
Below, we describe more detailed findings related to performance as a function of issue and potential error causes.
\subsubsection{Performance by accessibility \& safety issue}
Unsurprisingly, RASSAR's technical performance varies across accessibility and safety issue. In \Cref{tab:performancebyissue}, we show individual performance metrics for 13 commonly encountered issues. Almost all exhibit strong detection performance, with F1 scores exceeding 0.65. Some issues, like `\textit{bed height}' and `\textit{cabinet height}', even achieved 1.0 accuracy. However, there is one notable exception: `\textit{counter height}'. This particular issue, whose rubric requires a counter surface height of between 28 and 34 inches, shows subpar performance, a result that can be attributed to the RoomPlan API consistently classifying `\textit{kitchen counters}' as `\textit{storage units}.' This systematic error causes RASSAR to ignore `\textit{counter height}' in most scans.

\begin{table*}[]
\caption{Detection performance of different accessibility/safety issues. GT stands for ground truth from manual auditing. Prec is the abbreviation of precision. F1 is short for F1 score. Acc is the abbreviation of accuracy.}
\label{tab:performancebyissue}
\renewcommand{\arraystretch}{1.05}
\resizebox{0.99\textwidth}{!}{
\begin{tabular}{l|lrrrrrrrr}
\hline
\textbf{Category}                                                                            & \textbf{Issue Name}                                        & \multicolumn{1}{l}{\textbf{\begin{tabular}[c]{@{}l@{}}Count\\ of GT\end{tabular}}} & \multicolumn{1}{l}{\textbf{\begin{tabular}[c]{@{}l@{}}Count\\ of TP\end{tabular}}} & \multicolumn{1}{l}{\textbf{\begin{tabular}[c]{@{}l@{}}Count\\ of FP\end{tabular}}} & \multicolumn{1}{l}{\textbf{\begin{tabular}[c]{@{}l@{}}Count\\ of FN\end{tabular}}} & \multicolumn{1}{l}{\textbf{Precision}}              & \multicolumn{1}{l}{\textbf{Recall}}                 & \multicolumn{1}{l}{\textbf{F1 Score}}               & \multicolumn{1}{l}{\textbf{Accuracy}}               \\ \hline
                                                                                             & \cellcolor[HTML]{EFEFEF}Counter Height                     & \cellcolor[HTML]{EFEFEF}42                                                         & \cellcolor[HTML]{EFEFEF}3                                                          & \cellcolor[HTML]{EFEFEF}1                                                          & \cellcolor[HTML]{EFEFEF}39                                                         & \cellcolor[HTML]{EFEFEF}0.75                        & \cellcolor[HTML]{EFEFEF}0.07                        & \cellcolor[HTML]{EFEFEF}0.13                        & \cellcolor[HTML]{EFEFEF}0.07                        \\
                                                                                             & table height                                               & 42                                                                                 & 38                                                                                 & 8                                                                                  & 4                                                                                  & 0.83                                                & 0.91                                                & 0.86                                                & 0.76                                                \\
                                                                                             & \cellcolor[HTML]{EFEFEF}{\color[HTML]{000000} Door radius} & \cellcolor[HTML]{EFEFEF}{\color[HTML]{000000} 27}                                  & \cellcolor[HTML]{EFEFEF}{\color[HTML]{000000} 17}                                  & \cellcolor[HTML]{EFEFEF}{\color[HTML]{000000} 3}                                   & \cellcolor[HTML]{EFEFEF}{\color[HTML]{000000} 10}                                  & \cellcolor[HTML]{EFEFEF}{\color[HTML]{000000} 0.85} & \cellcolor[HTML]{EFEFEF}{\color[HTML]{000000} 0.63} & \cellcolor[HTML]{EFEFEF}{\color[HTML]{000000} 0.72} & \cellcolor[HTML]{EFEFEF}{\color[HTML]{000000} 0.57} \\
\multirow{-4}{*}{Object Dimension}                                                           & Bed Height                                                 & 15                                                                                 & 15                                                                                 & 0                                                                                  & 0                                                                                  & 1.00                                                & 1.00                                                & 1.00                                                & 1.00                                                \\ \cline{1-1}
                                                                                             & \cellcolor[HTML]{EFEFEF}Sink height                        & \cellcolor[HTML]{EFEFEF}57                                                         & \cellcolor[HTML]{EFEFEF}54                                                         & \cellcolor[HTML]{EFEFEF}0                                                          & \cellcolor[HTML]{EFEFEF}3                                                          & \cellcolor[HTML]{EFEFEF}1.00                        & \cellcolor[HTML]{EFEFEF}0.95                        & \cellcolor[HTML]{EFEFEF}0.97                        & \cellcolor[HTML]{EFEFEF}0.95                        \\
                                                                                             & Cabinet Height                                             & 48                                                                                 & 48                                                                                 & 0                                                                                  & 0                                                                                  & 1.00                                                & 1.00                                                & 1.00                                                & 1.00                                                \\
\multirow{-3}{*}{Object Position}                                                            & \cellcolor[HTML]{EFEFEF}Grab bar height                    & \cellcolor[HTML]{EFEFEF}15                                                         & \cellcolor[HTML]{EFEFEF}12                                                         & \cellcolor[HTML]{EFEFEF}8                                                          & \cellcolor[HTML]{EFEFEF}3                                                          & \cellcolor[HTML]{EFEFEF}0.60                        & \cellcolor[HTML]{EFEFEF}0.80                        & \cellcolor[HTML]{EFEFEF}0.69                        & \cellcolor[HTML]{EFEFEF}0.52                        \\ \cline{1-1}
                                                                                             & Rug                                                        & 48                                                                                 & 48                                                                                 & 2                                                                                  & 0                                                                                  & 0.96                                                & 1.00                                                & 0.98                                                & 0.96                                                \\
                                                                                             & \cellcolor[HTML]{EFEFEF}Medication                         & \cellcolor[HTML]{EFEFEF}21                                                         & \cellcolor[HTML]{EFEFEF}19                                                         & \cellcolor[HTML]{EFEFEF}18                                                         & \cellcolor[HTML]{EFEFEF}2                                                          & \cellcolor[HTML]{EFEFEF}0.51                        & \cellcolor[HTML]{EFEFEF}0.91                        & \cellcolor[HTML]{EFEFEF}0.66                        & \cellcolor[HTML]{EFEFEF}0.49                        \\
                                                                                             & Knife                                                      & 15                                                                                 & 14                                                                                 & 7                                                                                  & 1                                                                                  & 0.67                                                & 0.93                                                & 0.78                                                & 0.64                                                \\
\multirow{-4}{*}{\begin{tabular}[c]{@{}l@{}}Existence of \\ Risky Item\end{tabular}}         & \cellcolor[HTML]{EFEFEF}Scissors                           & \cellcolor[HTML]{EFEFEF}15                                                         & \cellcolor[HTML]{EFEFEF}15                                                         & \cellcolor[HTML]{EFEFEF}0                                                          & \cellcolor[HTML]{EFEFEF}0                                                          & \cellcolor[HTML]{EFEFEF}1.00                        & \cellcolor[HTML]{EFEFEF}1.00                        & \cellcolor[HTML]{EFEFEF}1.00                        & \cellcolor[HTML]{EFEFEF}1.00                        \\ \cline{1-1}
                                                                                             & No Grab bar near toilet                                    & 27                                                                                 & 24                                                                                 & 0                                                                                  & 3                                                                                  & 1.00                                                & 0.89                                                & 0.94                                                & 0.89                                                \\
\multirow{-2}{*}{\begin{tabular}[c]{@{}l@{}}Non-existence \\ of Assistive Item\end{tabular}} & \cellcolor[HTML]{EFEFEF}No Grab bar near tub               & \cellcolor[HTML]{EFEFEF}18                                                         & \cellcolor[HTML]{EFEFEF}16                                                         & \cellcolor[HTML]{EFEFEF}0                                                          & \cellcolor[HTML]{EFEFEF}2                                                          & \cellcolor[HTML]{EFEFEF}1.00                        & \cellcolor[HTML]{EFEFEF}0.89                        & \cellcolor[HTML]{EFEFEF}0.94                        & \cellcolor[HTML]{EFEFEF}0.89                        \\ \hline
\end{tabular}}
\end{table*}

\subsubsection{Causes of error}
\label{subsubsec:cause of error}
To better understand RASSAR errors, we classified them into six categories (\Cref{tab:errors}).
\begin{table*}[htb!]
\centering
\caption{The causes for RASSAR scan errors}
  \label{tab:errors}
\begin{tabular}{llll}
\hline
\textbf{Cause Name}                                                   & \textbf{Description}                                                                                                                 & \textbf{\begin{tabular}[c]{@{}l@{}}Total \\ count\end{tabular}} & \textbf{\begin{tabular}[c]{@{}l@{}}\% across\\ all errors\end{tabular}} \\ \hline
\begin{tabular}[c]{@{}l@{}}RoomPlan \\ Misclassification\end{tabular} & RoomPlan API misclassified an object                                                                                                 & 42                                                              & 32.8                                                                    \\
\rowcolor[HTML]{EFEFEF} 
\begin{tabular}[c]{@{}l@{}}RoomPlan \\ Measurement\end{tabular}       & \begin{tabular}[c]{@{}l@{}}RoomPlan API provided inaccurate object \\ measurement\end{tabular}                                       & 13                                                              & 10.1                                                                    \\
RoomPlan Miss                                                         & RoomPlan API missed an object                                                                                                        & 11                                                              & 8.6                                                                     \\
\rowcolor[HTML]{EFEFEF} 
\begin{tabular}[c]{@{}l@{}}YOLO \\ Misclassification\end{tabular}     & \begin{tabular}[c]{@{}l@{}}YOLO misclassified an object into another class, \\ or falsely report an non-existing object\end{tabular} & 55                                                              & 43.0                                                                    \\
YOLO Miss                                                             & YOLO missed an object of interest                                                                                                    & 3                                                               & 2.3                                                                     \\
\rowcolor[HTML]{EFEFEF} 
Raycast Issue                                                         & \begin{tabular}[c]{@{}l@{}}Raycasting process went wrong, resulting in\\ missing or misplacement of object\end{tabular}              & 4                                                               & 3.1                                                                     \\ \hline
\end{tabular}
\end{table*}
About half of all errors come from the object detection model, YoloV5, due to misclassifications (\textit{i.e.,} the YOLO model reported incorrect category labels for detected objects). Because indoor spaces contain many similar objects, misclassification is a significant technical challenge. For example, condiment bottles were misclassified as medicine, drawer handles were perceived as grab bars, and bed sheets were classified as rugs, leading to invalid results. These errors usually cause false positives, which can be alleviated by manual checking and filtering, as discussed in \Cref{subsubsec:visualization in AR}.

The other primary cause of error was due to RoomPlan API limitations, particularly  RoomPlan misclassifications. The RoomPlan API provided inaccurate object category labels, which led RASSAR to overlook underlying accessibility issues related to that object. For example, RoomPlan classifying a kitchen counter as storage, causing RASSAR to ignore underlying counter height issues.

\rev{\section{Study 3: User Study}
\label{sec:user study}}

\rev{
Finally, for our last evaluation, we performed a user study with six participants across our stakeholder groups, including two wheelchair users, one older adult, two new parents, an OT, and a person with severely low vision (legally blind)---see \autoref{tab:userstudy}. Among the wheelchair users, P1 used a manual chair and P4 an electric wheelchair—both had high-functioning upper-body motor control. Because we also envision RASSAR being used by non-stakeholders (\textit{e.g.,} Airbnb hosts), we also recruited one additional participant. For recruitment, we used mailing lists, outreach to local disability groups, and snowball sampling.}

\begin{figure}[h]
    
  \centering
  \includegraphics[width=1\linewidth]{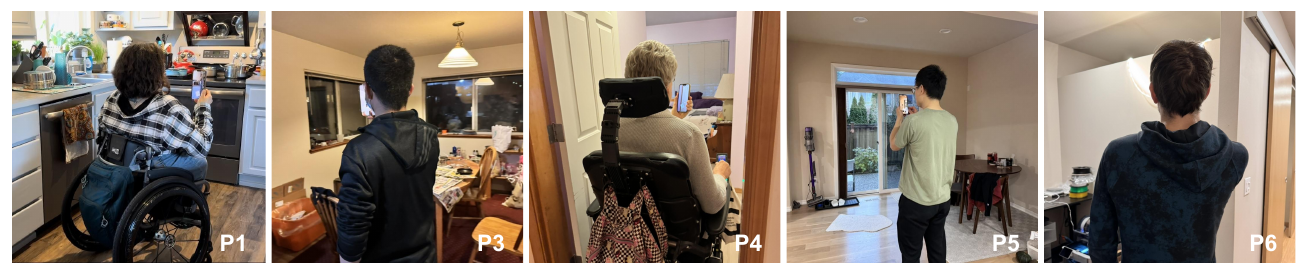}
  \caption{Study 3 participants conducting RASSAR scans.}
\label{fig:user scan}
\Description[Five photos, each showing a person conduct RASSAR scan.]{Five photos of study 3 participants conducting room scan with RASSAR. Each shows user holding a phone and conduct RASSAR scan.}
\end{figure}

\rev{\subsection{Procedure}
For the user study, two researchers visited participants' homes to conduct a three-part investigation: first, one researcher manually conducted a ground truth inspection (similar to Study 2) in selected rooms (\textit{e.g.,} kitchen, living room, and bathroom). Second, participants were provided with an iPhone 13 Pro Max with RASSAR installed. After a brief tutorial, they were asked to independently conduct a RASSAR scan of the same rooms. The research team observed the scanning behavior and took notes. Third and finally, the research team conducted a semi-structure debrief interview about RASSAR, inquiring about the overall experience, ideas for future work, and concluding with 7-point Likert scale ratings for usability, usefulness, perceived accuracy, and willingness to use RASSAR in the future (see \Cref{tab:userstudy}). For analysis, we compared ground truth to RASSAR's detections (similar to \Cref{study2procedure}) and thematically analyzed the interview data.}


\rev{\subsection{Findings}
Overall, all six participants could independently use and scan their space with RASSAR, rating the app highly usable and accurate (\textit{avg}=5.5 and 5.8 out of 7, respectively). Below, we expand on the technical performance, usability, usefulness, and user suggested improvements and application areas. }

\begin{table*}[]
\centering
    \caption{User study demographics, scan stats and user ratings.}
     \label{tab:userstudy}
\renewcommand{\arraystretch}{1.05}
\resizebox{0.99\textwidth}{!}{
\begin{tabular}{@{}lllrr|rrrrr|rrrr@{}}
\toprule
\multicolumn{1}{c}{}                                          & \multicolumn{1}{c}{}                                                                                           & \multicolumn{1}{c}{}                                                                                & \multicolumn{1}{c}{}                                                                                      & \multicolumn{1}{c|}{}                                                                                    & \multicolumn{1}{c}{}                                                                                              & \multicolumn{1}{c}{}                                                                                      & \multicolumn{1}{c}{}                                 & \multicolumn{1}{c}{}                                & \multicolumn{1}{c|}{}                               & \multicolumn{4}{c}{\textbf{Ratings from 1-7}}                                                                                                                                                                                                                                                                                                                                      \\ \cmidrule(l){11-14} 
\multicolumn{1}{c}{\multirow{-2}{*}{\textbf{ID}}} & \multicolumn{1}{c}{\multirow{-2}{*}{\textbf{\begin{tabular}[c]{@{}c@{}}Stakeholder\\  Identity\end{tabular}}}} & \multicolumn{1}{c}{\multirow{-2}{*}{\textbf{\begin{tabular}[c]{@{}c@{}}Home \\ Type\end{tabular}}}} & \multicolumn{1}{c}{\multirow{-2}{*}{\textbf{\begin{tabular}[c]{@{}c@{}}Scan Area\\  (sqm)\end{tabular}}}} & \multicolumn{1}{c|}{\multirow{-2}{*}{\textbf{\begin{tabular}[c]{@{}c@{}}Rooms \\ Scanned\end{tabular}}}} & \multicolumn{1}{c}{\multirow{-2}{*}{\textbf{\begin{tabular}[c]{@{}c@{}}Manual Audit \\ Time (Min)\end{tabular}}}} & \multicolumn{1}{c}{\multirow{-2}{*}{\textbf{\begin{tabular}[c]{@{}c@{}}Scan Time\\  (Min)\end{tabular}}}} & \multicolumn{1}{c}{\multirow{-2}{*}{\textbf{Prec.}}} & \multicolumn{1}{c}{\multirow{-2}{*}{\textbf{Rec.}}} & \multicolumn{1}{c|}{\multirow{-2}{*}{\textbf{Acc}}} & \multicolumn{1}{l}{\textbf{\begin{tabular}[c]{@{}l@{}}App \\ usability\end{tabular}}} & \multicolumn{1}{l}{\textbf{\begin{tabular}[c]{@{}l@{}}Usefulness\\  of results\end{tabular}}} & \multicolumn{1}{l}{\textbf{\begin{tabular}[c]{@{}l@{}}Detection \\ performance\end{tabular}}} & \multicolumn{1}{l}{\textbf{\begin{tabular}[c]{@{}l@{}}Willingness\\  to use\end{tabular}}} \\ \midrule
P1                                                            & Wheelchair user                                                                                                & House                                                                                               & 120                                                                                                      & 4                                                                                                        & 12                                                                                                                & 5.02                                                                                                      & 0.74                                                 & 0.67                                                & 0.54                                                & \cellcolor[HTML]{65A255}{\color[HTML]{000000} 6}                                      & \cellcolor[HTML]{ABAFAB}{\color[HTML]{000000} 4}                                              & \cellcolor[HTML]{65A255}{\color[HTML]{000000} 6}                                              & \cellcolor[HTML]{429818}{\color[HTML]{000000} 7}                                           \\
\rowcolor[HTML]{EFEFEF} 
P2                                                            & OT                                                                                                             & Apartment                                                                                           & 43                                                                                                       & 2                                                                                                        & 21                                                                                                                & 3.08                                                                                                      & 0.54                                                 & 0.88                                                & 0.5                                                 & \cellcolor[HTML]{87A982}{\color[HTML]{000000} 5}                                      & \cellcolor[HTML]{ABAFAB}{\color[HTML]{000000} 4}                                              & \cellcolor[HTML]{87A982}{\color[HTML]{000000} 5}                                              & \cellcolor[HTML]{FF1110}{\color[HTML]{000000} 1}                                           \\
P3                                                            &         None                                                                                                       & House                                                                                               & 60                                                                                                       & 2                                                                                                        & 14                                                                                                                & 2.25                                                                                                      & 0.92                                                 & 0.85                                                & 0.79                                                & \cellcolor[HTML]{87A982}{\color[HTML]{000000} 5}                                      & \cellcolor[HTML]{65A255}{\color[HTML]{000000} 6}                                              & \cellcolor[HTML]{65A255}{\color[HTML]{000000} 6}                                              & \cellcolor[HTML]{ABAFAB}{\color[HTML]{000000} 4}                                           \\
\rowcolor[HTML]{EFEFEF} 
P4                                                            & \begin{tabular}[c]{@{}l@{}}Wheelchair user, \\ older adult\end{tabular}                                        & Apartment                                                                                           & 70                                                                                                       & 4                                                                                                        & 10                                                                                                                & 3.25                                                                                                      & 1                                                    & 0.73                                                & 0.73                                                & \cellcolor[HTML]{65A255}{\color[HTML]{000000} 6}                                      & \cellcolor[HTML]{65A255}{\color[HTML]{000000} 6}                                              & \cellcolor[HTML]{429818}{\color[HTML]{000000} 7}                                              & \cellcolor[HTML]{65A255}{\color[HTML]{000000} 6}                                           \\
P5                                                            & New parents                                                                                                    & House                                                                                               & 150                                                                                                      & 3                                                                                                        & 7                                                                                                                 & 3.38                                                                                                      & 0.67                                                 & 0.46                                                & 0.38                                                & \cellcolor[HTML]{65A255}{\color[HTML]{000000} 6}                                      &      -                                                                                         & \cellcolor[HTML]{87A982}{\color[HTML]{000000} 5}                                              & \cellcolor[HTML]{87A982}{\color[HTML]{000000} 5}                                           \\
\rowcolor[HTML]{EFEFEF} 
{\color[HTML]{000000} P6}                                     & {\color[HTML]{000000} BLV}                                                                                     & {\color[HTML]{000000} Apartment}                                                                    & {\color[HTML]{000000} 55}                                                                                & {\color[HTML]{000000} 3}                                                                                 & {\color[HTML]{000000} 7}                                                                                          & {\color[HTML]{000000} 3.05}                                                                               & {\color[HTML]{000000} 0.88}                          & {\color[HTML]{000000} 0.78}                         & {\color[HTML]{000000} 0.7}                          & \cellcolor[HTML]{87A982}{\color[HTML]{000000} 5}                                      & \cellcolor[HTML]{FF1110}{\color[HTML]{000000} 1}                                              & \cellcolor[HTML]{65A255}{\color[HTML]{000000} 6}                                              & \cellcolor[HTML]{65A255}{\color[HTML]{000000} 6}                                           \\ \midrule
\multicolumn{5}{r|}{\textbf{Average}}                                                                                                                                                                                                                                                                                                                                                                                                                                                                       & \textbf{11.83}                                                                                                    & \textbf{3.34}                                                                                             & \textbf{0.79}                                        & \textbf{0.73}                                       & \textbf{0.61}                                       & \textbf{5.5}                                                                          & \textbf{4.2}                                                                                  & \textbf{5.83}                                                                                 & \textbf{4.83}                                                                              \\ \bottomrule
\end{tabular}}
\end{table*}

\rev{\textbf{Technical Performance}
 The scans resulted in an average precision and recall of 0.79 and 0.73, which is slightly lower than the scanning performance in Study 2, where precision was 0.86 and recall was 0.83. This difference can likely be attributed to Study 2's scans being performed by a single member of our research team, whereas in the current study, actual stakeholder users conducted the scans. When asked to rate their perception of RASSAR's detection performance, the average response from participants was also high: 5.8/7 (\textit{min:} 5). ``\textit{The 3D reconstruction results are way better than my imagination}'' (P5). Regarding scanning speed, participants completed their RASSAR-based scans in 3.3 minutes on average—3.5x faster than the ground truth manual auditing (\textit{avg=}11.8min). }

\rev{\textbf{Usability.} As noted above, all six participants successfully finished their home scans with RASSAR and rated the app highly usable (\textit{avg=5.6/7; min: 5}). We observed different scanning practices, which may have also impacted technical performance. For example, P3, P4 and P6 constantly tilted phone up and down during scan, while P2 and P5 fixed phone vertical thus missed accessibility and safety issues on the floor. This oversight may be alleviated by visual or audio hints for missed indoor surfaces during scan. For P6, the low vision user, he was able to complete his scan with the audio assistance feature. 
Interestingly, though unexpected, other participants also found utility in the audio assistance. For example, both P4 and P5, who are sighted, conducted RASSAR scans with audio feedback enabled, which they felt increased their awareness of the scan progress and detected issues. ``\textit{The audio helped me a lot in understanding what's happening and what should I focus on during the scan.}'' (P5)}

\rev{\textbf{Usefulness.} In terms of perceived usefulness, the results were more nuanced: P3 and P4 rated usefulness as 6/7 because ``\textit{I would never know that the throw rug could be an issue}'' (P3) and ``\textit{These results will be potentially useful if I want to pick a new home}'' (P4). However, P1 and P2 rated usefulness as 4/7. P1, a manual chair user, felt that the ADA-based issues did not match her needs since ``\textit{My house is very usable to me, but in RASSAR's head, it wasn't}''. Interestingly, P1 still rated the app highly usable and wanted to use the app in the future. For the OT (P2), they wanted RASSAR to be more customizable (``\textit{it's not customized enough}'')—a feature we hope to support in the future. P6, who is low vision, provided high ratings for all aspects except usefulness. He found the detected issues interesting and helpful, but rated usefulness low because the last part of the scan process---the scan summary of results---was not as accessible as he needed.}

\rev{\textbf{Potential applications.} In the debrief interviews, participants offered a variety of use cases for RASSAR. P1, who rated full marks for her willingness to use RASSAR, envisions RASSAR as a way to scan, assess, and share the accessibility of public spaces: ``\textit{It would be really helpful in public spaces. I'm on several Facebook groups for people with disability. And we will all turn to write reviews of different places that we go. [With RASSAR], we can have a more objective way to share that information.}'' Similarly, P2 (the OT), felt that RASSAR could be used to help raise awareness about accessibility issues and scale better than on-site OT inspections: ``\textit{What OT can do [to help people] is very limited. There are so many people in need but very few have access to OT service.}''. Although P2 rated low for her own willingness to use, she was happy to recommend RASSAR to others. ``\textit{I might not use it since I already have this knowledge. But I would recommend it to my family or anyone in need. It's so convenient and way better than me lecturing them!}''}

\rev{P3 found more personal use cases instead. ``\textit{If I move to a new home, or when I get injured someday, I might want to use this (to check for home safety)}''. Similarly, P4 thinks RASSAR can be used in auditing new homes to help make the most suitable purchase. ``\textit{It would be helpful for auditing another place.}'' P5 thinks RASSAR could be helpful for child-proofing if more children-related issues get implemented in the future.}

\rev{\textbf{System Improvements.}
Participants provided valuable feedback on enhancing user interaction with the system. A prevalent set of suggestions regarded the summary view (\autoref{fig:summary}), particularly in improving the manipulation of the 3D model (P1) and organizing scan results by specific rooms, such as the kitchen, bedroom, and restroom (P2). Additionally, both P2 and P5 wanted the scan summary in a list format. P6 also echoed this suggestion since a list view is more compatible with screen readers. Another set of recommendations revolved around improving scanning support. P5 suggested additional audio or visual cues to alert users when they miss certain indoor surfaces, like the floor. Similarly, P6 wanted extra audio alerts for the proximity of indoor objects, in order to prevent BLV users from bumping into barriers. Furthermore, P3 advocated for more direct visual guidance, such as arrows, to assist users in conducting thorough scans without overlooking key areas. }



\section{Discussion}
We introduce the first mobile AR system to detect accessibility and safety issues in home spaces. Using RASSAR, individuals can semi-automatically audit their home spaces and generate \rev{real-time} 3D reconstructions that highlight accessibility and safety issues. \rev{Additionally, we conducted three studies to examine real-world needs, technical performance, and user experience of our proposed method. Below we elaborate on the implications of our findings and opportunities for future work.}




\subsection{Application Scenarios}
Unlike prior work in automatic accessibility auditing---which often involves complex data collection processes \cite{balado_automatic_2017,fu_human-centric_2020} or specific hardware \cite{serna_urban_2013,ayala-alfaro_automatic_2021}---the RASSAR system is a plug-and-play application on smartphones. This ease of use significantly expands its appeal and potential applications. From the formative study (\Cref{subsection:formativeresults}) \rev{and user study (\Cref{sec:user study})}, we identified numerous possible use-cases for RASSAR, which we expand on below.

\textbf{Prior-visit Auditing.}
One common challenge identified in the formative study is the uncertainty about a location's accessibility before visiting, including rental spaces, such as hotel rooms or Airbnb accommodations, \rev{public spaces, like restaurants and shops,} as well as friends' \rev{or family's }homes. \rev{Based on our user study feedback (\Cref{sec:user study}),} RASSAR emerges as a practical solution by offering a standardized scanning and detection process for previewing accessibility and safety issues. Site owners could employ RASSAR to ensure that their spaces meet \rev{general} accessibility requirements \rev{or share the scan results to help people preview and evaluate a space before visiting}. \rev{Future work should examine sharing interfaces to support this desired feature.}

\textbf{Improve Spaces to Accommodate Life Changes.}
All lives undergo transformation, which can impact accessibility and safety. 
\rev{For people undergoing significant life events such as illness, childbirth, or the need to care for older family members, we found that RASSAR can serve as a convenient evaluation tool that raises awareness of potential risks under such changes, and could also provide home renovation suggestions such as removal of dangerous items and adjustment of object dimensions.} 

\textbf{Complement OT's Home Visits.}
As previous research \cite{renda_feasibility_2018} and our own studies with OTs indicate, it is often challenging for individuals to request, schedule, and pay for OT home visits. With RASSAR, OTs can offer remote assistance by reviewing the system's scanning results, which not only encompass visual information but also include 3D positions and measurements. Consequently, remote auditing processes could become more standardized, efficient, and precise compared to existing methods like video calls.

\textbf{Beyond residential.} While RASSAR is presently designed for residential spaces, its technical framework could be applied to non-residential areas such as offices, schools, and restaurants. In initial work, we conducted successful tests of RASSAR in two offices. In the future, we would like to incorporate new rubrics for non-residential spaces and examine methods to upload and view assessments.

\rev{
\subsection{Detection Performance}
As described in \autoref{subsec:evalresults} and \autoref{sec:user study}, RASSAR yielded an average precision/recall of 0.86/0.83 when operated by our research team, and 0.79/0.73 by stakeholder participants. Compared with manual auditing, scans are also about 3.5x faster. While preliminary, these results are promising. Still, RASSAR's performance could be improved. Most error cases were caused by deficiencies in our YOLOV5 model (\Cref{subsubsec:object detection}) and the RoomPlan API. In the future, we plan to further improve RASSAR performance by expanding our object detection model to also detect furniture categories to correct RoomPlan misclassifications, and also expand training dataset on micro indoor objects to improve detection performance. More work is necessary to determine what accuracy is required for the different application scenarios proposed above.}

\rev{\subsection{Accessibility Issue Scope}}

Currently, RASSAR detects 20 types of accessibility and safety issues across four categories (\Cref{fig:rubrics}). We plan to expand this list based on needs found in the literature \rev{ (\textit{e.g.} detecting sharp edges of furniture~\cite{niemeyer1994g94})}, our formative study findings (\textit{e.g.,} home entrances, stairs and bath facilities), \rev{ and user study findings (\textit{e.g. }wheelchair maneuvering spaces).}

\rev{
\subsection{Potential Beyond the ADA}
The current RASSAR system relies on ADA design guideline as the main source of accessibility rubrics. But as found in both our formative and summative user studies, ADA design guidelines are often minimum requirements and may not fit everyone's needs. ``\textit{ADA is just a fixed standard for reference, thus it cannot ensure fitting on everyone.}'' (P2, Study 3 )  Similarly, one wheelchair participant from Study 3 complained that she once found out a non-ADA hotel room also worked great for her when the ADA units were booked out. In this case, the binary classification of "ADA accessible" became exclusionary. With RASSAR's custom JSON-based rubric definitions, assessments could be personalized to individual needs. Future work should explore rubric authoring interfaces.}

\section{Conclusion}
We introduced RASSAR, a mobile AR system that semi-automatically audits indoor residential spaces for accessibility and safety issues. Built with state-of-the-art mobile LiDAR scanners and mobile computer vision models, RASSAR efficiently and effectively identifies, localizes, and visually displays accessibility and safety issues and provides recommendations for mitigation. RASSAR is both customizable, letting users specify their target accessibility communities, and verifiable, letting users manually verify detected issues. Our technical evaluation (Study 2) in ten home spaces shows that \textit{RASSAR}’s performance is accurate, consistent, and efficient. \rev{Our initial user study (Study 3) further demonstrates RASSAR's potential among stakeholder groups}. Our work advances the literature on indoor accessibility and safety auditing, contributes indoor accessibility object detection model and dataset to the research community, while simultaneously opening up new research avenues for human-AI collaborative indoor auditing.

\begin{acks}
This research was supported by Meta, the UW Reality Lab, NSF award \#1834629, an NSF GRFP, and the Center for Research and Education on Accessible Technology and Experiences (CREATE). We thank our interview and user study participants for their participation. 
\end{acks}

\bibliographystyle{ACM-Reference-Format}
\bibliography{ref}


\begin{thebibliography}{64}


\ifx \showCODEN    \undefined \def \showCODEN     #1{\unskip}     \fi
\ifx \showDOI      \undefined \def \showDOI       #1{#1}\fi
\ifx \showISBNx    \undefined \def \showISBNx     #1{\unskip}     \fi
\ifx \showISBNxiii \undefined \def \showISBNxiii  #1{\unskip}     \fi
\ifx \showISSN     \undefined \def \showISSN      #1{\unskip}     \fi
\ifx \showLCCN     \undefined \def \showLCCN      #1{\unskip}     \fi
\ifx \shownote     \undefined \def \shownote      #1{#1}          \fi
\ifx \showarticletitle \undefined \def \showarticletitle #1{#1}   \fi
\ifx \showURL      \undefined \def \showURL       {\relax}        \fi
\providecommand\bibfield[2]{#2}
\providecommand\bibinfo[2]{#2}
\providecommand\natexlab[1]{#1}
\providecommand\showeprint[2][]{arXiv:#2}

\bibitem[~(2010)]%
        {noauthor_ada_2010}
\bibfield{author}{\bibinfo{person}{Department of~Justice  }.} \bibinfo{year}{2010}\natexlab{}.
\newblock \bibinfo{title}{{ADA} {Standards} for {Accessible} {Design}}.
\newblock
\newblock
\urldef\tempurl%
\url{https://www.ada.gov/law-and-regs/design-standards/}
\showURL{%
\tempurl}


\bibitem[Anjanappa(2022)]%
        {anjanappa_deep_2022}
\bibfield{author}{\bibinfo{person}{Geethanjali Anjanappa}.} \bibinfo{year}{2022}\natexlab{}.
\newblock \bibinfo{title}{Deep learning on {3D} point clouds for safety-related asset management in buildings}.
\newblock
\newblock
\urldef\tempurl%
\url{http://essay.utwente.nl/91463/}
\showURL{%
\tempurl}
\newblock
\shownote{Publisher: University of Twente}.


\bibitem[Apple({[n.\,d.]})]%
        {apple_core_nodate}
\bibfield{author}{\bibinfo{person}{Apple}.} \bibinfo{year}{[n.\,d.]}\natexlab{}.
\newblock \bibinfo{title}{Core {ML}}.
\newblock
\newblock
\urldef\tempurl%
\url{https://developer.apple.com/documentation/coreml}
\showURL{%
\tempurl}


\bibitem[Apple(2020)]%
        {apple_apple_2020}
\bibfield{author}{\bibinfo{person}{Apple}.} \bibinfo{year}{2020}\natexlab{}.
\newblock \bibinfo{title}{Apple introduces {iPhone} 12 {Pro} and {iPhone} 12 {Pro} {Max} with {5G}}.
\newblock
\newblock
\urldef\tempurl%
\url{https://www.apple.com/newsroom/2020/10/apple-introduces-iphone-12-pro-and-iphone-12-pro-max-with-5g/}
\showURL{%
\tempurl}


\bibitem[Apple(2022)]%
        {apple_roomplan_2022}
\bibfield{author}{\bibinfo{person}{Apple}.} \bibinfo{year}{2022}\natexlab{}.
\newblock \bibinfo{title}{{RoomPlan} - {Augmented} {Reality}}.
\newblock
\newblock
\urldef\tempurl%
\url{https://developer.apple.com/augmented-reality/roomplan/}
\showURL{%
\tempurl}


\bibitem[Apple(2023)]%
        {apple_raycast_2023}
\bibfield{author}{\bibinfo{person}{Apple}.} \bibinfo{year}{2023}\natexlab{}.
\newblock \bibinfo{title}{Raycast}.
\newblock
\newblock
\urldef\tempurl%
\url{https://developer.apple.com/documentation/arkit/arsession/3132065-raycast}
\showURL{%
\tempurl}


\bibitem[Ayala-Alfaro et~al\mbox{.}(2021)]%
        {ayala-alfaro_automatic_2021}
\bibfield{author}{\bibinfo{person}{V. Ayala-Alfaro}, \bibinfo{person}{J.~A. Vilchis-Mar}, \bibinfo{person}{F.~E. Correa-Tome}, {and} \bibinfo{person}{J.~P. Ramirez-Paredes}.} \bibinfo{year}{2021}\natexlab{}.
\newblock \bibinfo{title}{Automatic {Mapping} with {Obstacle} {Identification} for {Indoor} {Human} {Mobility} {Assessment}}.
\newblock
\newblock
\urldef\tempurl%
\url{https://doi.org/10.48550/arXiv.2111.12690}
\showDOI{\tempurl}
\newblock
\shownote{arXiv:2111.12690 [cs]}.


\bibitem[Balado et~al\mbox{.}(2017)]%
        {balado_automatic_2017}
\bibfield{author}{\bibinfo{person}{J. Balado}, \bibinfo{person}{L. Díaz-Vilariño}, \bibinfo{person}{P. Arias}, {and} \bibinfo{person}{M. Soilán}.} \bibinfo{year}{2017}\natexlab{}.
\newblock \showarticletitle{Automatic building accessibility diagnosis from point clouds}.
\newblock \bibinfo{journal}{\emph{Automation in Construction}}  \bibinfo{volume}{82} (\bibinfo{date}{Oct.} \bibinfo{year}{2017}), \bibinfo{pages}{103--111}.
\newblock
\showISSN{0926-5805}
\urldef\tempurl%
\url{https://doi.org/10.1016/j.autcon.2017.06.026}
\showDOI{\tempurl}


\bibitem[Braun and Clarke(2019)]%
        {braun_reflecting_2019}
\bibfield{author}{\bibinfo{person}{Virginia Braun} {and} \bibinfo{person}{Victoria Clarke}.} \bibinfo{year}{2019}\natexlab{}.
\newblock \showarticletitle{Reflecting on reflexive thematic analysis}.
\newblock \bibinfo{journal}{\emph{Qualitative Research in Sport, Exercise and Health}} \bibinfo{volume}{11}, \bibinfo{number}{4} (\bibinfo{date}{Aug.} \bibinfo{year}{2019}), \bibinfo{pages}{589--597}.
\newblock
\showISSN{2159-676X}
\urldef\tempurl%
\url{https://doi.org/10.1080/2159676X.2019.1628806}
\showDOI{\tempurl}
\newblock
\shownote{Publisher: Routledge \_eprint: https://doi.org/10.1080/2159676X.2019.1628806}.


\bibitem[Chiu and Oliver(2006)]%
        {chiu2006factor}
\bibfield{author}{\bibinfo{person}{Teresa Chiu} {and} \bibinfo{person}{Rosemary Oliver}.} \bibinfo{year}{2006}\natexlab{}.
\newblock \showarticletitle{Factor analysis and construct validity of the SAFER-HOME}.
\newblock \bibinfo{journal}{\emph{OTJR: Occupation, Participation and Health}} \bibinfo{volume}{26}, \bibinfo{number}{4} (\bibinfo{year}{2006}), \bibinfo{pages}{132--142}.
\newblock


\bibitem[Clemson(1997)]%
        {clemson_home_1997}
\bibfield{author}{\bibinfo{person}{Lindy Clemson}.} \bibinfo{year}{1997}\natexlab{}.
\newblock \bibinfo{booktitle}{\emph{Home {Fall} {Hazards}: {A} {Guide} to {Identifying} {Fall} {Hazards} in the {Homes} of {Elderly} {People} and an {Accompaniment} to the {Assessment} {Tool}, {The} {Westmead} {Home} {Safety} {Assessment} ({WeHSA})}}.
\newblock \bibinfo{publisher}{Co-ordinates Publications}.
\newblock


\bibitem[Dao and Thill(2018)]%
        {dao_three-dimensional_2018}
\bibfield{author}{\bibinfo{person}{Thi Hong~Diep Dao} {and} \bibinfo{person}{Jean-Claude Thill}.} \bibinfo{year}{2018}\natexlab{}.
\newblock \showarticletitle{Three-dimensional indoor network accessibility auditing for floor plan design}.
\newblock \bibinfo{journal}{\emph{Transactions in GIS}} \bibinfo{volume}{22}, \bibinfo{number}{1} (\bibinfo{year}{2018}), \bibinfo{pages}{288--310}.
\newblock
\showISSN{1467-9671}
\urldef\tempurl%
\url{https://doi.org/10.1111/tgis.12310}
\showDOI{\tempurl}
\newblock
\shownote{\_eprint: https://onlinelibrary.wiley.com/doi/pdf/10.1111/tgis.12310}.


\bibitem[Deng et~al\mbox{.}(2022)]%
        {deng_semantic_2022}
\bibfield{author}{\bibinfo{person}{Hui Deng}, \bibinfo{person}{Mao Tian}, \bibinfo{person}{Zhibin Ou}, {and} \bibinfo{person}{Yichuan Deng}.} \bibinfo{year}{2022}\natexlab{}.
\newblock \showarticletitle{A semantic framework for on-site evacuation routing based on awareness of obstacle accessibility}.
\newblock \bibinfo{journal}{\emph{Automation in Construction}}  \bibinfo{volume}{136} (\bibinfo{date}{April} \bibinfo{year}{2022}), \bibinfo{pages}{104154}.
\newblock
\showISSN{0926-5805}
\urldef\tempurl%
\url{https://doi.org/10.1016/j.autcon.2022.104154}
\showDOI{\tempurl}


\bibitem[Díaz-Vilariño et~al\mbox{.}(2019)]%
        {diaz-vilarino_obstacle-aware_2019}
\bibfield{author}{\bibinfo{person}{Lucía Díaz-Vilariño}, \bibinfo{person}{Pawel Boguslawski}, \bibinfo{person}{Kourosh Khoshelham}, {and} \bibinfo{person}{Henrique Lorenzo}.} \bibinfo{year}{2019}\natexlab{}.
\newblock \showarticletitle{Obstacle-{Aware} {Indoor} {Pathfinding} {Using} {Point} {Clouds}}.
\newblock \bibinfo{journal}{\emph{ISPRS International Journal of Geo-Information}} \bibinfo{volume}{8}, \bibinfo{number}{5} (\bibinfo{date}{May} \bibinfo{year}{2019}), \bibinfo{pages}{233}.
\newblock
\showISSN{2220-9964}
\urldef\tempurl%
\url{https://doi.org/10.3390/ijgi8050233}
\showDOI{\tempurl}
\newblock
\shownote{Number: 5 Publisher: Multidisciplinary Digital Publishing Institute}.


\bibitem[Díaz~Vilariño et~al\mbox{.}(2022)]%
        {diaz_vilarino_3d_2022}
\bibfield{author}{\bibinfo{person}{Lucia Díaz~Vilariño}, \bibinfo{person}{Ha Tran}, \bibinfo{person}{Ernesto Frías}, \bibinfo{person}{Jesus Balado~Frias}, {and} \bibinfo{person}{Kourosh Khoshelham}.} \bibinfo{year}{2022}\natexlab{}.
\newblock \showarticletitle{{3D} {MAPPING} {OF} {INDOOR} {AND} {OUTDOOR} {ENVIRONMENTS} {USING} {APPLE} {SMART} {DEVICES}}.
\newblock \bibinfo{journal}{\emph{The International Archives of the Photogrammetry, Remote Sensing and Spatial Information Sciences}}  \bibinfo{volume}{XLIII-B4-2022} (\bibinfo{date}{June} \bibinfo{year}{2022}), \bibinfo{pages}{303--308}.
\newblock
\urldef\tempurl%
\url{https://doi.org/10.5194/isprs-archives-XLIII-B4-2022-303-2022}
\showDOI{\tempurl}


\bibitem[Froehlich et~al\mbox{.}(2012)]%
        {froehlich2012design}
\bibfield{author}{\bibinfo{person}{Jon Froehlich}, \bibinfo{person}{Leah Findlater}, \bibinfo{person}{Marilyn Ostergren}, \bibinfo{person}{Solai Ramanathan}, \bibinfo{person}{Josh Peterson}, \bibinfo{person}{Inness Wragg}, \bibinfo{person}{Eric Larson}, \bibinfo{person}{Fabia Fu}, \bibinfo{person}{Mazhengmin Bai}, \bibinfo{person}{Shwetak Patel}, {et~al\mbox{.}}} \bibinfo{year}{2012}\natexlab{}.
\newblock \showarticletitle{The design and evaluation of prototype eco-feedback displays for fixture-level water usage data}. In \bibinfo{booktitle}{\emph{Proceedings of the SIGCHI conference on human factors in computing systems}}. \bibinfo{pages}{2367--2376}.
\newblock


\bibitem[Fu et~al\mbox{.}(2020)]%
        {fu_human-centric_2020}
\bibfield{author}{\bibinfo{person}{Qiang Fu}, \bibinfo{person}{Hongbo Fu}, \bibinfo{person}{Hai Yan}, \bibinfo{person}{Bin Zhou}, \bibinfo{person}{Xiaowu Chen}, {and} \bibinfo{person}{Xueming Li}.} \bibinfo{year}{2020}\natexlab{}.
\newblock \showarticletitle{Human-centric metrics for indoor scene assessment and synthesis}.
\newblock \bibinfo{journal}{\emph{Graphical Models}}  \bibinfo{volume}{110} (\bibinfo{date}{July} \bibinfo{year}{2020}), \bibinfo{pages}{101073}.
\newblock
\showISSN{1524-0703}
\urldef\tempurl%
\url{https://doi.org/10.1016/j.gmod.2020.101073}
\showDOI{\tempurl}


\bibitem[Fänge and Iwarsson(1999)]%
        {fange_physical_1999}
\bibfield{author}{\bibinfo{person}{Agneta Fänge} {and} \bibinfo{person}{Susannne Iwarsson}.} \bibinfo{year}{1999}\natexlab{}.
\newblock \showarticletitle{Physical {Housing} {Environment}: {Development} of a {Self}-{Assessment} {Instrument}}.
\newblock \bibinfo{journal}{\emph{Canadian Journal of Occupational Therapy}} \bibinfo{volume}{66}, \bibinfo{number}{5} (\bibinfo{date}{Dec.} \bibinfo{year}{1999}), \bibinfo{pages}{250--260}.
\newblock
\showISSN{0008-4174}
\urldef\tempurl%
\url{https://doi.org/10.1177/000841749906600507}
\showDOI{\tempurl}
\newblock
\shownote{Publisher: SAGE Publications Inc}.


\bibitem[Fänge and Iwarsson(2005)]%
        {fange_changes_2005}
\bibfield{author}{\bibinfo{person}{Agneta Fänge} {and} \bibinfo{person}{Susanne Iwarsson}.} \bibinfo{year}{2005}\natexlab{}.
\newblock \showarticletitle{Changes in accessibility and usability in housing: an exploration of the housing adaptation process}.
\newblock \bibinfo{journal}{\emph{Occupational Therapy International}} \bibinfo{volume}{12}, \bibinfo{number}{1} (\bibinfo{year}{2005}), \bibinfo{pages}{44--59}.
\newblock
\showISSN{1557-0703}
\urldef\tempurl%
\url{https://doi.org/10.1002/oti.14}
\showDOI{\tempurl}
\newblock
\shownote{\_eprint: https://onlinelibrary.wiley.com/doi/pdf/10.1002/oti.14}.


\bibitem[Gitlin et~al\mbox{.}(2002)]%
        {gitlin_evaluating_2002}
\bibfield{author}{\bibinfo{person}{L.~N. Gitlin}, \bibinfo{person}{S. Schinfeld}, \bibinfo{person}{L. Winter}, \bibinfo{person}{M. Corcoran}, \bibinfo{person}{A.~A. Boyce}, {and} \bibinfo{person}{W. Hauck}.} \bibinfo{year}{2002}\natexlab{}.
\newblock \showarticletitle{Evaluating home environments of persons with dementia: interrater reliability and validity of the {Home} {Environmental} {Assessment} {Protocol} ({HEAP})}.
\newblock \bibinfo{journal}{\emph{Disability and Rehabilitation}} \bibinfo{volume}{24}, \bibinfo{number}{1-3} (\bibinfo{date}{Jan.} \bibinfo{year}{2002}), \bibinfo{pages}{59--71}.
\newblock
\showISSN{0963-8288, 1464-5165}
\urldef\tempurl%
\url{https://doi.org/10.1080/09638280110066325}
\showDOI{\tempurl}


\bibitem[Gray et~al\mbox{.}(2008)]%
        {gray_subjective_2008}
\bibfield{author}{\bibinfo{person}{David~B. Gray}, \bibinfo{person}{Holly~H. Hollingsworth}, \bibinfo{person}{Susan Stark}, {and} \bibinfo{person}{Kerri~A. Morgan}.} \bibinfo{year}{2008}\natexlab{}.
\newblock \showarticletitle{A subjective measure of environmental facilitators and barriers to participation for people with mobility limitations}.
\newblock \bibinfo{journal}{\emph{Disability and Rehabilitation}} \bibinfo{volume}{30}, \bibinfo{number}{6} (\bibinfo{date}{Jan.} \bibinfo{year}{2008}), \bibinfo{pages}{434--457}.
\newblock
\showISSN{0963-8288, 1464-5165}
\urldef\tempurl%
\url{https://doi.org/10.1080/09638280701625377}
\showDOI{\tempurl}


\bibitem[Hara et~al\mbox{.}(2016)]%
        {hara2016design}
\bibfield{author}{\bibinfo{person}{Kotaro Hara}, \bibinfo{person}{Christine Chan}, {and} \bibinfo{person}{Jon~E Froehlich}.} \bibinfo{year}{2016}\natexlab{}.
\newblock \showarticletitle{The design of assistive location-based technologies for people with ambulatory disabilities: A formative study}. In \bibinfo{booktitle}{\emph{Proceedings of the 2016 CHI conference on human factors in computing systems}}. \bibinfo{pages}{1757--1768}.
\newblock


\bibitem[Hashemi and Karimi(2016)]%
        {hashemi_indoor_2016}
\bibfield{author}{\bibinfo{person}{Mahdi Hashemi} {and} \bibinfo{person}{Hassan~A. Karimi}.} \bibinfo{year}{2016}\natexlab{}.
\newblock \showarticletitle{Indoor {Spatial} {Model} and {Accessibility} {Index} for {Emergency} {Evacuation} of {People} with {Disabilities}}.
\newblock \bibinfo{journal}{\emph{Journal of Computing in Civil Engineering}} \bibinfo{volume}{30}, \bibinfo{number}{4} (\bibinfo{date}{July} \bibinfo{year}{2016}), \bibinfo{pages}{04015056}.
\newblock
\showISSN{1943-5487}
\urldef\tempurl%
\url{https://doi.org/10.1061/(ASCE)CP.1943-5487.0000534}
\showDOI{\tempurl}
\newblock
\shownote{Publisher: American Society of Civil Engineers}.


\bibitem[Horowitz et~al\mbox{.}(2016)]%
        {horowitz_use_2016}
\bibfield{author}{\bibinfo{person}{Beverly~P. Horowitz}, \bibinfo{person}{Tiffany Almonte}, {and} \bibinfo{person}{Andrea Vasil}.} \bibinfo{year}{2016}\natexlab{}.
\newblock \showarticletitle{Use of the {Home} {Safety} {Self}-{Assessment} {Tool} ({HSSAT}) within {Community} {Health} {Education} to {Improve} {Home} {Safety}}.
\newblock \bibinfo{journal}{\emph{Occupational Therapy In Health Care}} \bibinfo{volume}{30}, \bibinfo{number}{4} (\bibinfo{date}{Oct.} \bibinfo{year}{2016}), \bibinfo{pages}{356--372}.
\newblock
\showISSN{0738-0577}
\urldef\tempurl%
\url{https://doi.org/10.1080/07380577.2016.1191695}
\showDOI{\tempurl}
\newblock
\shownote{Publisher: Taylor \& Francis \_eprint: https://doi.org/10.1080/07380577.2016.1191695}.


\bibitem[Horvath et~al\mbox{.}(2013)]%
        {horvath_clinical_2013}
\bibfield{author}{\bibinfo{person}{Kathy~J. Horvath}, \bibinfo{person}{Scott~A. Trudeau}, \bibinfo{person}{James~L. Rudolph}, \bibinfo{person}{Paulette~A. Trudeau}, \bibinfo{person}{Mary~E. Duffy}, {and} \bibinfo{person}{Dan Berlowitz}.} \bibinfo{year}{2013}\natexlab{}.
\newblock \showarticletitle{Clinical {Trial} of a {Home} {Safety} {Toolkit} for {Alzheimer}’s {Disease}}.
\newblock \bibinfo{journal}{\emph{International Journal of Alzheimer’s Disease}}  \bibinfo{volume}{2013} (\bibinfo{date}{Sept.} \bibinfo{year}{2013}), \bibinfo{pages}{e913606}.
\newblock
\showISSN{2090-8024}
\urldef\tempurl%
\url{https://doi.org/10.1155/2013/913606}
\showDOI{\tempurl}
\newblock
\shownote{Publisher: Hindawi}.


\bibitem[Housing and inc(1996)]%
        {housing_fair_1996}
\bibfield{author}{\bibinfo{person}{United States Department of Housing {and} Urban Development Office~of Housing} {and} \bibinfo{person}{Barrier Free~Environments inc}.} \bibinfo{year}{1996}\natexlab{}.
\newblock \bibinfo{booktitle}{\emph{Fair {Housing} {Act} {Design} {Manual}: {A} {Manual} to {Assist} {Designers} and {Builders} in {Meeting} the {Accessibility} {Requirements} of the {Fair} {Housing} {Act}}}.
\newblock \bibinfo{publisher}{U.S. Department of Housing and Urban Development, Office of Fair Housing and Equal Opportunity and the Office of Housing}.
\newblock


\bibitem[Hughes(2021)]%
        {hughes2021krippendorffsalpha}
\bibfield{author}{\bibinfo{person}{John Hughes}.} \bibinfo{year}{2021}\natexlab{}.
\newblock \showarticletitle{krippendorffsalpha: An R package for measuring agreement using Krippendorff's alpha coefficient}.
\newblock \bibinfo{journal}{\emph{arXiv preprint arXiv:2103.12170}} (\bibinfo{year}{2021}).
\newblock


\bibitem[Imrie(2003)]%
        {imrie2003housing}
\bibfield{author}{\bibinfo{person}{Rob Imrie}.} \bibinfo{year}{2003}\natexlab{}.
\newblock \showarticletitle{Housing quality and the provision of accessible homes}.
\newblock \bibinfo{journal}{\emph{Housing Studies}} \bibinfo{volume}{18}, \bibinfo{number}{3} (\bibinfo{year}{2003}), \bibinfo{pages}{387--408}.
\newblock


\bibitem[Ito and Biljecki(2021)]%
        {ito_assessing_2021}
\bibfield{author}{\bibinfo{person}{Koichi Ito} {and} \bibinfo{person}{Filip Biljecki}.} \bibinfo{year}{2021}\natexlab{}.
\newblock \showarticletitle{Assessing bikeability with street view imagery and computer vision}.
\newblock \bibinfo{journal}{\emph{Transportation Research Part C: Emerging Technologies}}  \bibinfo{volume}{132} (\bibinfo{date}{Nov.} \bibinfo{year}{2021}), \bibinfo{pages}{103371}.
\newblock
\showISSN{0968-090X}
\urldef\tempurl%
\url{https://doi.org/10.1016/j.trc.2021.103371}
\showDOI{\tempurl}


\bibitem[Iwarsson(1999)]%
        {iwarsson_housing_1999}
\bibfield{author}{\bibinfo{person}{Susanne Iwarsson}.} \bibinfo{year}{1999}\natexlab{}.
\newblock \showarticletitle{The {Housing} {Enabler}: {An} {Objective} {Tool} for {Assessing} {Accessibility}}.
\newblock \bibinfo{journal}{\emph{British Journal of Occupational Therapy}} \bibinfo{volume}{62}, \bibinfo{number}{11} (\bibinfo{date}{Nov.} \bibinfo{year}{1999}), \bibinfo{pages}{491--497}.
\newblock
\showISSN{0308-0226}
\urldef\tempurl%
\url{https://doi.org/10.1177/030802269906201104}
\showDOI{\tempurl}
\newblock
\shownote{Publisher: SAGE Publications Ltd STM}.


\bibitem[Jain et~al\mbox{.}(2015)]%
        {jain_head-mounted_2015}
\bibfield{author}{\bibinfo{person}{Dhruv Jain}, \bibinfo{person}{Leah Findlater}, \bibinfo{person}{Jamie Gilkeson}, \bibinfo{person}{Benjamin Holland}, \bibinfo{person}{Ramani Duraiswami}, \bibinfo{person}{Dmitry Zotkin}, \bibinfo{person}{Christian Vogler}, {and} \bibinfo{person}{Jon~E. Froehlich}.} \bibinfo{year}{2015}\natexlab{}.
\newblock \showarticletitle{Head-{Mounted} {Display} {Visualizations} to {Support} {Sound} {Awareness} for the {Deaf} and {Hard} of {Hearing}}. In \bibinfo{booktitle}{\emph{Proceedings of the 33rd {Annual} {ACM} {Conference} on {Human} {Factors} in {Computing} {Systems}}} \emph{(\bibinfo{series}{{CHI} '15})}. \bibinfo{publisher}{Association for Computing Machinery}, \bibinfo{address}{New York, NY, USA}, \bibinfo{pages}{241--250}.
\newblock
\showISBNx{978-1-4503-3145-6}
\urldef\tempurl%
\url{https://doi.org/10.1145/2702123.2702393}
\showDOI{\tempurl}


\bibitem[Jocher et~al\mbox{.}(2022)]%
        {jocher_ultralyticsyolov5_2022}
\bibfield{author}{\bibinfo{person}{Glenn Jocher}, \bibinfo{person}{Ayush Chaurasia}, \bibinfo{person}{Alex Stoken}, \bibinfo{person}{Jirka Borovec}, \bibinfo{person}{NanoCode012}, \bibinfo{person}{Yonghye Kwon}, \bibinfo{person}{Kalen Michael}, \bibinfo{person}{TaoXie}, \bibinfo{person}{Jiacong Fang}, \bibinfo{person}{imyhxy}, \bibinfo{person}{Lorna}, \bibinfo{person}{Zeng Yifu}, \bibinfo{person}{Colin Wong}, \bibinfo{person}{Abhiram V}, \bibinfo{person}{Diego Montes}, \bibinfo{person}{Zhiqiang Wang}, \bibinfo{person}{Cristi Fati}, \bibinfo{person}{Jebastin Nadar}, \bibinfo{person}{Laughing}, \bibinfo{person}{UnglvKitDe}, \bibinfo{person}{Victor Sonck}, \bibinfo{person}{tkianai}, \bibinfo{person}{yxNONG}, \bibinfo{person}{Piotr Skalski}, \bibinfo{person}{Adam Hogan}, \bibinfo{person}{Dhruv Nair}, \bibinfo{person}{Max Strobel}, {and} \bibinfo{person}{Mrinal Jain}.} \bibinfo{year}{2022}\natexlab{}.
\newblock \bibinfo{title}{ultralytics/yolov5: v7.0 - {YOLOv5} {SOTA} {Realtime} {Instance} {Segmentation}}.
\newblock
\newblock
\urldef\tempurl%
\url{https://doi.org/10.5281/zenodo.7347926}
\showDOI{\tempurl}


\bibitem[{Kids Clinic}(2023)]%
        {KidsClinicSafety}
\bibfield{author}{\bibinfo{person}{{Kids Clinic}}.} \bibinfo{year}{2023}\natexlab{}.
\newblock \bibinfo{title}{Home Safety Checklist}.
\newblock
\newblock
\urldef\tempurl%
\url{https://kidsclinic.pediatricweb.com/Medical-Content/Safety/Home-Safety-Checklist}
\showURL{%
\tempurl}
\newblock
\shownote{[Online; accessed 8-December-2023]}.


\bibitem[Krippendorff(2011)]%
        {krippendorff2011computing}
\bibfield{author}{\bibinfo{person}{Klaus Krippendorff}.} \bibinfo{year}{2011}\natexlab{}.
\newblock \showarticletitle{Computing Krippendorff's alpha-reliability}.
\newblock  (\bibinfo{year}{2011}).
\newblock


\bibitem[Kuznetsova et~al\mbox{.}(2020)]%
        {OpenImages}
\bibfield{author}{\bibinfo{person}{Alina Kuznetsova}, \bibinfo{person}{Hassan Rom}, \bibinfo{person}{Neil Alldrin}, \bibinfo{person}{Jasper Uijlings}, \bibinfo{person}{Ivan Krasin}, \bibinfo{person}{Jordi Pont-Tuset}, \bibinfo{person}{Shahab Kamali}, \bibinfo{person}{Stefan Popov}, \bibinfo{person}{Matteo Malloci}, \bibinfo{person}{Alexander Kolesnikov}, \bibinfo{person}{Tom Duerig}, {and} \bibinfo{person}{Vittorio Ferrari}.} \bibinfo{year}{2020}\natexlab{}.
\newblock \showarticletitle{The Open Images Dataset V4: Unified image classification, object detection, and visual relationship detection at scale}.
\newblock \bibinfo{journal}{\emph{IJCV}} (\bibinfo{year}{2020}).
\newblock


\bibitem[Lange et~al\mbox{.}(2021)]%
        {lange_strategically_2021}
\bibfield{author}{\bibinfo{person}{Marvin Lange}, \bibinfo{person}{Reuben Kirkham}, {and} \bibinfo{person}{Benjamin Tannert}.} \bibinfo{year}{2021}\natexlab{}.
\newblock \showarticletitle{Strategically {Using} {Applied} {Machine} {Learning} for {Accessibility} {Documentation} in the {Built} {Environment}}. In \bibinfo{booktitle}{\emph{Human-{Computer} {Interaction} – {INTERACT} 2021}} \emph{(\bibinfo{series}{Lecture {Notes} in {Computer} {Science}})}, \bibfield{editor}{\bibinfo{person}{Carmelo Ardito}, \bibinfo{person}{Rosa Lanzilotti}, \bibinfo{person}{Alessio Malizia}, \bibinfo{person}{Helen Petrie}, \bibinfo{person}{Antonio Piccinno}, \bibinfo{person}{Giuseppe Desolda}, {and} \bibinfo{person}{Kori Inkpen}} (Eds.). \bibinfo{publisher}{Springer International Publishing}, \bibinfo{address}{Cham}, \bibinfo{pages}{426--448}.
\newblock
\showISBNx{978-3-030-85616-8}
\urldef\tempurl%
\url{https://doi.org/10.1007/978-3-030-85616-8_25}
\showDOI{\tempurl}


\bibitem[Law et~al\mbox{.}(1996)]%
        {law1996person}
\bibfield{author}{\bibinfo{person}{Mary Law}, \bibinfo{person}{Barbara Cooper}, \bibinfo{person}{Susan Strong}, \bibinfo{person}{Debra Stewart}, \bibinfo{person}{Patricia Rigby}, {and} \bibinfo{person}{Lori Letts}.} \bibinfo{year}{1996}\natexlab{}.
\newblock \showarticletitle{The person-environment-occupation model: A transactive approach to occupational performance}.
\newblock \bibinfo{journal}{\emph{Canadian journal of occupational therapy}} \bibinfo{volume}{63}, \bibinfo{number}{1} (\bibinfo{year}{1996}), \bibinfo{pages}{9--23}.
\newblock


\bibitem[Luetzenburg et~al\mbox{.}(2021)]%
        {luetzenburg_evaluation_2021}
\bibfield{author}{\bibinfo{person}{Gregor Luetzenburg}, \bibinfo{person}{Aart Kroon}, {and} \bibinfo{person}{Anders~A. Bjørk}.} \bibinfo{year}{2021}\natexlab{}.
\newblock \showarticletitle{Evaluation of the {Apple} {iPhone} 12 {Pro} {LiDAR} for an {Application} in {Geosciences}}.
\newblock \bibinfo{journal}{\emph{Scientific Reports}} \bibinfo{volume}{11}, \bibinfo{number}{1} (\bibinfo{date}{Nov.} \bibinfo{year}{2021}), \bibinfo{pages}{22221}.
\newblock
\showISSN{2045-2322}
\urldef\tempurl%
\url{https://doi.org/10.1038/s41598-021-01763-9}
\showDOI{\tempurl}
\newblock
\shownote{Number: 1 Publisher: Nature Publishing Group}.


\bibitem[Luttrell(2022)]%
        {luttrell_data_2022}
\bibfield{author}{\bibinfo{person}{Joseph Luttrell}.} \bibinfo{year}{2022}\natexlab{}.
\newblock \showarticletitle{Data {Collection} and {Machine} {Learning} {Methods} for {Automated} {Pedestrian} {Facility} {Detection} and {Mensuration}}.
\newblock \bibinfo{journal}{\emph{Dissertations}} (\bibinfo{date}{Aug.} \bibinfo{year}{2022}).
\newblock
\urldef\tempurl%
\url{https://aquila.usm.edu/dissertations/2034}
\showURL{%
\tempurl}


\bibitem[Microsoft({[n.\,d.]})]%
        {Microsoft}
\bibfield{author}{\bibinfo{person}{Microsoft}.} \bibinfo{year}{[n.\,d.]}\natexlab{}.
\newblock \bibinfo{title}{Bing image search API: Microsoft bing}.
\newblock
\newblock
\urldef\tempurl%
\url{https://www.microsoft.com/en-us/bing/apis/bing-image-search-api}
\showURL{%
\tempurl}


\bibitem[Nguyen et~al\mbox{.}(2021)]%
        {nguyen2021reducing}
\bibfield{author}{\bibinfo{person}{Thien Nguyen}, \bibinfo{person}{Elizabeth~M Combs}, \bibinfo{person}{Pamela~J Wright}, {and} \bibinfo{person}{Cynthia~F Corbett}.} \bibinfo{year}{2021}\natexlab{}.
\newblock \showarticletitle{Reducing fall risks among visually impaired older adults}.
\newblock \bibinfo{journal}{\emph{Home healthcare now}} \bibinfo{volume}{39}, \bibinfo{number}{4} (\bibinfo{year}{2021}), \bibinfo{pages}{186--193}.
\newblock


\bibitem[Niemeyer and Vogel(1994)]%
        {niemeyer1994g94}
\bibfield{author}{\bibinfo{person}{Shirley Niemeyer} {and} \bibinfo{person}{Michael~P Vogel}.} \bibinfo{year}{1994}\natexlab{}.
\newblock \showarticletitle{G94-1213 Child Care Environment Safety Checklist}.
\newblock \bibinfo{journal}{\emph{Historical Materials from University of Nebraska-Lincoln Extension}} (\bibinfo{year}{1994}), \bibinfo{pages}{1325}.
\newblock


\bibitem[Organization(2018)]%
        {world_health_organization_who_2018}
\bibfield{author}{\bibinfo{person}{World~Health Organization}.} \bibinfo{year}{2018}\natexlab{}.
\newblock \bibinfo{title}{{WHO} {Housing} and health guidelines}.
\newblock
\newblock
\urldef\tempurl%
\url{https://www.who.int/publications-detail-redirect/9789241550376}
\showURL{%
\tempurl}


\bibitem[Osborne et~al\mbox{.}(2008)]%
        {osborne_home_2008}
\bibfield{author}{\bibinfo{person}{Helen Osborne}, \bibinfo{person}{Tom Scott}, {and} \bibinfo{person}{United~Spinal Association}.} \bibinfo{year}{2008}\natexlab{}.
\newblock \showarticletitle{Home {Safety} for {People} {With} {Disabilities}}.
\newblock \bibinfo{journal}{\emph{in Motion}} \bibinfo{volume}{18}, \bibinfo{number}{5} (\bibinfo{year}{2008}).
\newblock
\urldef\tempurl%
\url{https://www.cdss.ca.gov/agedblinddisabled/res/VPTC2/5%20Injury%20and%20Fall%20Prevention/Home_Safety_for_People_with_Disabilities.pdf}
\showURL{%
\tempurl}


\bibitem[Powers(2020)]%
        {powers2020evaluation}
\bibfield{author}{\bibinfo{person}{David~MW Powers}.} \bibinfo{year}{2020}\natexlab{}.
\newblock \showarticletitle{Evaluation: from precision, recall and F-measure to ROC, informedness, markedness and correlation}.
\newblock \bibinfo{journal}{\emph{arXiv preprint arXiv:2010.16061}} (\bibinfo{year}{2020}).
\newblock


\bibitem[RENDA and LAPE(2018)]%
        {renda_feasibility_2018}
\bibfield{author}{\bibinfo{person}{MARNIE RENDA} {and} \bibinfo{person}{JENNIFER~E. LAPE}.} \bibinfo{year}{2018}\natexlab{}.
\newblock \showarticletitle{Feasibility and {Effectiveness} of {Telehealth} {Occupational} {Therapy} {Home} {Modification} {Interventions}}.
\newblock \bibinfo{journal}{\emph{International Journal of Telerehabilitation}} \bibinfo{volume}{10}, \bibinfo{number}{1} (\bibinfo{date}{Aug.} \bibinfo{year}{2018}), \bibinfo{pages}{3--14}.
\newblock
\showISSN{1945-2020}
\urldef\tempurl%
\url{https://doi.org/10.5195/ijt.2018.6244}
\showDOI{\tempurl}


\bibitem[Rev.com({[n.\,d.]})]%
        {revcom_transcribe_nodate}
\bibfield{author}{\bibinfo{person}{Rev.com}.} \bibinfo{year}{[n.\,d.]}\natexlab{}.
\newblock \bibinfo{title}{Transcribe {Speech} to {Text}}.
\newblock
\newblock
\urldef\tempurl%
\url{https://www.rev.com/}
\showURL{%
\tempurl}


\bibitem[Romero et~al\mbox{.}(2017)]%
        {romero_development_2017}
\bibfield{author}{\bibinfo{person}{Sergio Romero}, \bibinfo{person}{Mi Lee}, \bibinfo{person}{Ivana Simic}, \bibinfo{person}{Charles Levy}, {and} \bibinfo{person}{Jon Sanford}.} \bibinfo{year}{2017}\natexlab{}.
\newblock \showarticletitle{Development and validation of a remote home safety protocol}.
\newblock \bibinfo{journal}{\emph{Disability and Rehabilitation: Assistive Technology}}  \bibinfo{volume}{13} (\bibinfo{date}{March} \bibinfo{year}{2017}), \bibinfo{pages}{1--7}.
\newblock
\urldef\tempurl%
\url{https://doi.org/10.1080/17483107.2017.1300345}
\showDOI{\tempurl}


\bibitem[Saha et~al\mbox{.}(2022)]%
        {saha_visualizing_2022}
\bibfield{author}{\bibinfo{person}{Manaswi Saha}, \bibinfo{person}{Siddhant Patil}, \bibinfo{person}{Emily Cho}, \bibinfo{person}{Evie Yu-Yen Cheng}, \bibinfo{person}{Chris Horng}, \bibinfo{person}{Devanshi Chauhan}, \bibinfo{person}{Rachel Kangas}, \bibinfo{person}{Richard McGovern}, \bibinfo{person}{Anthony Li}, \bibinfo{person}{Jeffrey Heer}, {and} \bibinfo{person}{Jon~E. Froehlich}.} \bibinfo{year}{2022}\natexlab{}.
\newblock \showarticletitle{Visualizing {Urban} {Accessibility}: {Investigating} {Multi}-{Stakeholder} {Perspectives} through a {Map}-based {Design} {Probe} {Study}}. In \bibinfo{booktitle}{\emph{Proceedings of the 2022 {CHI} {Conference} on {Human} {Factors} in {Computing} {Systems}}} \emph{(\bibinfo{series}{{CHI} '22})}. \bibinfo{publisher}{Association for Computing Machinery}, \bibinfo{address}{New York, NY, USA}, \bibinfo{pages}{1--14}.
\newblock
\showISBNx{978-1-4503-9157-3}
\urldef\tempurl%
\url{https://doi.org/10.1145/3491102.3517460}
\showDOI{\tempurl}


\bibitem[Saha et~al\mbox{.}(2019)]%
        {saha_project_2019}
\bibfield{author}{\bibinfo{person}{Manaswi Saha}, \bibinfo{person}{Michael Saugstad}, \bibinfo{person}{Hanuma~Teja Maddali}, \bibinfo{person}{Aileen Zeng}, \bibinfo{person}{Ryan Holland}, \bibinfo{person}{Steven Bower}, \bibinfo{person}{Aditya Dash}, \bibinfo{person}{Sage Chen}, \bibinfo{person}{Anthony Li}, \bibinfo{person}{Kotaro Hara}, {and} \bibinfo{person}{Jon Froehlich}.} \bibinfo{year}{2019}\natexlab{}.
\newblock \showarticletitle{Project {Sidewalk}: {A} {Web}-based {Crowdsourcing} {Tool} for {Collecting} {Sidewalk} {Accessibility} {Data} {At} {Scale}}. In \bibinfo{booktitle}{\emph{Proceedings of the 2019 {CHI} {Conference} on {Human} {Factors} in {Computing} {Systems}}}. \bibinfo{publisher}{ACM}, \bibinfo{address}{Glasgow Scotland Uk}, \bibinfo{pages}{1--14}.
\newblock
\showISBNx{978-1-4503-5970-2}
\urldef\tempurl%
\url{https://doi.org/10.1145/3290605.3300292}
\showDOI{\tempurl}


\bibitem[Serna and Marcotegui(2013)]%
        {serna_urban_2013}
\bibfield{author}{\bibinfo{person}{Andrés Serna} {and} \bibinfo{person}{Beatriz Marcotegui}.} \bibinfo{year}{2013}\natexlab{}.
\newblock \showarticletitle{Urban accessibility diagnosis from mobile laser scanning data}.
\newblock \bibinfo{journal}{\emph{ISPRS Journal of Photogrammetry and Remote Sensing}}  \bibinfo{volume}{84} (\bibinfo{date}{Oct.} \bibinfo{year}{2013}), \bibinfo{pages}{23--32}.
\newblock
\showISSN{0924-2716}
\urldef\tempurl%
\url{https://doi.org/10.1016/j.isprsjprs.2013.07.001}
\showDOI{\tempurl}


\bibitem[Sharif et~al\mbox{.}(2021)]%
        {sharif_experimental_2021}
\bibfield{author}{\bibinfo{person}{Ather Sharif}, \bibinfo{person}{Paari Gopal}, \bibinfo{person}{Michael Saugstad}, \bibinfo{person}{Shiven Bhatt}, \bibinfo{person}{Raymond Fok}, \bibinfo{person}{Galen Weld}, \bibinfo{person}{Kavi Asher Mankoff~Dey}, {and} \bibinfo{person}{Jon E.~Froehlich}.} \bibinfo{year}{2021}\natexlab{}.
\newblock \showarticletitle{Experimental {Crowd}+{AI} {Approaches} to {Track} {Accessibility} {Features} in {Sidewalk} {Intersections} {Over} {Time}}. In \bibinfo{booktitle}{\emph{Proceedings of the 23rd {International} {ACM} {SIGACCESS} {Conference} on {Computers} and {Accessibility}}} \emph{(\bibinfo{series}{{ASSETS} '21})}. \bibinfo{publisher}{Association for Computing Machinery}, \bibinfo{address}{New York, NY, USA}, \bibinfo{pages}{1--5}.
\newblock
\showISBNx{978-1-4503-8306-6}
\urldef\tempurl%
\url{https://doi.org/10.1145/3441852.3476549}
\showDOI{\tempurl}


\bibitem[Smith et~al\mbox{.}(2008)]%
        {smith_aging_2008}
\bibfield{author}{\bibinfo{person}{Stanley~K. Smith}, \bibinfo{person}{Stefan Rayer}, {and} \bibinfo{person}{Eleanor~A. Smith}.} \bibinfo{year}{2008}\natexlab{}.
\newblock \showarticletitle{Aging and {Disability}: {Implications} for the {Housing} {Industry} and {Housing} {Policy} in the {United} {States}}.
\newblock \bibinfo{journal}{\emph{Journal of the American Planning Association}} \bibinfo{volume}{74}, \bibinfo{number}{3} (\bibinfo{date}{July} \bibinfo{year}{2008}), \bibinfo{pages}{289--306}.
\newblock
\showISSN{0194-4363}
\urldef\tempurl%
\url{https://doi.org/10.1080/01944360802197132}
\showDOI{\tempurl}
\newblock
\shownote{Publisher: Routledge \_eprint: https://doi.org/10.1080/01944360802197132}.


\bibitem[Steinfeld et~al\mbox{.}(1998)]%
        {steinfeld_home_1998}
\bibfield{author}{\bibinfo{person}{Edward Steinfeld}, \bibinfo{person}{Danise~R. Levine}, {and} \bibinfo{person}{Scott~M. Shea}.} \bibinfo{year}{1998}\natexlab{}.
\newblock \showarticletitle{Home modifications and the fair housing law}.
\newblock \bibinfo{journal}{\emph{Technology and Disability}} \bibinfo{volume}{8}, \bibinfo{number}{1-2} (\bibinfo{date}{Jan.} \bibinfo{year}{1998}), \bibinfo{pages}{15--35}.
\newblock
\showISSN{1055-4181}
\urldef\tempurl%
\url{https://doi.org/10.3233/TAD-1998-81-203}
\showDOI{\tempurl}
\newblock
\shownote{Publisher: IOS Press}.


\bibitem[Struckmeyer et~al\mbox{.}(2021)]%
        {struckmeyer_home_2021}
\bibfield{author}{\bibinfo{person}{Linda Struckmeyer}, \bibinfo{person}{Jane Morgan-Daniel}, \bibinfo{person}{Sherry Ahrentzen}, {and} \bibinfo{person}{Carlyn Ellison}.} \bibinfo{year}{2021}\natexlab{}.
\newblock \showarticletitle{Home {Modification} {Assessments} for {Accessibility} and {Aesthetics}: {A} {Rapid} {Review}}.
\newblock \bibinfo{journal}{\emph{HERD: Health Environments Research \& Design Journal}} \bibinfo{volume}{14}, \bibinfo{number}{2} (\bibinfo{date}{April} \bibinfo{year}{2021}), \bibinfo{pages}{313--327}.
\newblock
\showISSN{1937-5867, 2167-5112}
\urldef\tempurl%
\url{https://doi.org/10.1177/1937586720960704}
\showDOI{\tempurl}


\bibitem[Su et~al\mbox{.}(2022)]%
        {su_towards_2022}
\bibfield{author}{\bibinfo{person}{Xia Su}, \bibinfo{person}{Kaiming Cheng}, \bibinfo{person}{Han Zhang}, \bibinfo{person}{Jaewook Lee}, {and} \bibinfo{person}{Jon~E. Froehlich}.} \bibinfo{year}{2022}\natexlab{}.
\newblock \bibinfo{title}{Towards {Semi}-automatic {Detection} and {Localization} of {Indoor} {Accessibility} {Issues} using {Mobile} {Depth} {Scanning} and {Computer} {Vision}}.
\newblock
\newblock
\urldef\tempurl%
\url{https://doi.org/10.48550/arXiv.2210.02533}
\showDOI{\tempurl}
\newblock
\shownote{arXiv:2210.02533 [cs]}.


\bibitem[Swenor et~al\mbox{.}(2016)]%
        {swenor_evaluation_2016}
\bibfield{author}{\bibinfo{person}{Bonnielin~K. Swenor}, \bibinfo{person}{Andrea~V. Yonge}, \bibinfo{person}{Victoria Goldhammer}, \bibinfo{person}{Rhonda Miller}, \bibinfo{person}{Laura~N. Gitlin}, {and} \bibinfo{person}{Pradeep Ramulu}.} \bibinfo{year}{2016}\natexlab{}.
\newblock \showarticletitle{Evaluation of the {Home} {Environment} {Assessment} for the {Visually} {Impaired} ({HEAVI}): an instrument designed to quantify fall-related hazards in the visually impaired}.
\newblock \bibinfo{journal}{\emph{BMC Geriatrics}} \bibinfo{volume}{16}, \bibinfo{number}{1} (\bibinfo{date}{Dec.} \bibinfo{year}{2016}), \bibinfo{pages}{214}.
\newblock
\showISSN{1471-2318}
\urldef\tempurl%
\url{https://doi.org/10.1186/s12877-016-0391-2}
\showDOI{\tempurl}


\bibitem[Taira and Carlson(2014)]%
        {taira_aging_2014}
\bibfield{author}{\bibinfo{person}{Ellen~D. Taira} {and} \bibinfo{person}{Jodi Carlson}.} \bibinfo{year}{2014}\natexlab{}.
\newblock \bibinfo{booktitle}{\emph{Aging in {Place}: {Designing}, {Adapting}, and {Enhancing} the {Home} {Environment}}}.
\newblock \bibinfo{publisher}{Routledge}, \bibinfo{address}{New York}.
\newblock
\showISBNx{978-1-315-82152-8}
\urldef\tempurl%
\url{https://doi.org/10.4324/9781315821528}
\showDOI{\tempurl}


\bibitem[Together({[n.\,d.]})]%
        {rebuilding_together_safe_nodate}
\bibfield{author}{\bibinfo{person}{Rebuilding Together}.} \bibinfo{year}{[n.\,d.]}\natexlab{}.
\newblock \bibinfo{title}{Safe at {Home} {Checklist}}.
\newblock
\newblock


\bibitem[Tomita et~al\mbox{.}(2014)]%
        {tomita_psychometrics_2014}
\bibfield{author}{\bibinfo{person}{Machiko~R. Tomita}, \bibinfo{person}{Sumandeep Saharan}, \bibinfo{person}{Sheela Rajendran}, \bibinfo{person}{Susan~M. Nochajski}, {and} \bibinfo{person}{Jo~A. Schweitzer}.} \bibinfo{year}{2014}\natexlab{}.
\newblock \showarticletitle{Psychometrics of the {Home} {Safety} {Self}-{Assessment} {Tool} ({HSSAT}) to {Prevent} {Falls} in {Community}-{Dwelling} {Older} {Adults}}.
\newblock \bibinfo{journal}{\emph{The American Journal of Occupational Therapy}} \bibinfo{volume}{68}, \bibinfo{number}{6} (\bibinfo{date}{Nov.} \bibinfo{year}{2014}), \bibinfo{pages}{711--718}.
\newblock
\showISSN{0272-9490}
\urldef\tempurl%
\url{https://doi.org/10.5014/ajot.2014.010801}
\showDOI{\tempurl}


\bibitem[Ultralytics(2023)]%
        {yolov5github}
\bibfield{author}{\bibinfo{person}{Ultralytics}.} \bibinfo{year}{Accessed September 11, 2023}\natexlab{}.
\newblock \bibinfo{title}{YOLOv5 GitHub Repository}.
\newblock \bibinfo{howpublished}{\url{https://github.com/ultralytics/yolov5}}.
\newblock


\bibitem[Vogt et~al\mbox{.}(2021)]%
        {vogt_comparison_2021}
\bibfield{author}{\bibinfo{person}{Maximilian Vogt}, \bibinfo{person}{Adrian Rips}, {and} \bibinfo{person}{Claus Emmelmann}.} \bibinfo{year}{2021}\natexlab{}.
\newblock \showarticletitle{Comparison of {iPad} {Pro}®’s {LiDAR} and {TrueDepth} {Capabilities} with an {Industrial} {3D} {Scanning} {Solution}}.
\newblock \bibinfo{journal}{\emph{Technologies}} \bibinfo{volume}{9}, \bibinfo{number}{2} (\bibinfo{date}{June} \bibinfo{year}{2021}), \bibinfo{pages}{25}.
\newblock
\showISSN{2227-7080}
\urldef\tempurl%
\url{https://doi.org/10.3390/technologies9020025}
\showDOI{\tempurl}
\newblock
\shownote{Number: 2 Publisher: Multidisciplinary Digital Publishing Institute}.


\bibitem[Weld et~al\mbox{.}(2019)]%
        {weld_deep_2019}
\bibfield{author}{\bibinfo{person}{Galen Weld}, \bibinfo{person}{Esther Jang}, \bibinfo{person}{Anthony Li}, \bibinfo{person}{Aileen Zeng}, \bibinfo{person}{Kurtis Heimerl}, {and} \bibinfo{person}{Jon~E. Froehlich}.} \bibinfo{year}{2019}\natexlab{}.
\newblock \showarticletitle{Deep {Learning} for {Automatically} {Detecting} {Sidewalk} {Accessibility} {Problems} {Using} {Streetscape} {Imagery}}. In \bibinfo{booktitle}{\emph{The 21st {International} {ACM} {SIGACCESS} {Conference} on {Computers} and {Accessibility}}}. \bibinfo{publisher}{ACM}, \bibinfo{address}{Pittsburgh PA USA}, \bibinfo{pages}{196--209}.
\newblock
\showISBNx{978-1-4503-6676-2}
\urldef\tempurl%
\url{https://doi.org/10.1145/3308561.3353798}
\showDOI{\tempurl}


\bibitem[Yang et~al\mbox{.}(2020)]%
        {9201064}
\bibfield{author}{\bibinfo{person}{Xingbin Yang}, \bibinfo{person}{Liyang Zhou}, \bibinfo{person}{Hanqing Jiang}, \bibinfo{person}{Zhongliang Tang}, \bibinfo{person}{Yuanbo Wang}, \bibinfo{person}{Hujun Bao}, {and} \bibinfo{person}{Guofeng Zhang}.} \bibinfo{year}{2020}\natexlab{}.
\newblock \showarticletitle{Mobile3DRecon: Real-time Monocular 3D Reconstruction on a Mobile Phone}.
\newblock \bibinfo{journal}{\emph{IEEE Transactions on Visualization and Computer Graphics}} \bibinfo{volume}{26}, \bibinfo{number}{12} (\bibinfo{year}{2020}), \bibinfo{pages}{3446--3456}.
\newblock
\urldef\tempurl%
\url{https://doi.org/10.1109/TVCG.2020.3023634}
\showDOI{\tempurl}


\end{thebibliography}

\appendix

\newpage


\section{JSON Example}
\label{appendix:json}

\lstdefinestyle{json-counter}{
    basicstyle=\ttfamily,
    breaklines=true,
    breakanywhere=true,
    showstringspaces=true,
    frame=lines,
    language=json,
    literate=
      *{":}{{{\color{red}\bfseries:}}}{1}
      {,}{{{\color{red}\bfseries,}}}{1}
      {\{}{{{\color{red}\bfseries\{}}}{1}
      {\}}{{{\color{red}\bfseries\}}}}{1}
      {[}{{{\color{red}\bfseries[}}}{1}      
      {]}{{{\color{red}\bfseries]}}}{1}
      }

\begin{minted}[linenos=true,breaklines=true,breakanywhere=true]{json}
"Counter":{
    "Dim_Height":{
        "Community":["Wheelchair"],
        "Dependency":null,
        "Dimension":{
            "Comparison":"Between",
            "Value":[28,34]
        },
        "RelativePosition":{
            "Comparison":null,
            "Value":null
        },
    "Existence":null,
    "Note":"replace PLACEHOLDER to either 'short' or 'tall' depends on the actual height of the counter.",
    "Message":"Warning: Counter is too PLACEHOLDER.",
    "Description":"According to ADA compliance, counters must be at the proper height (this often is 28-34 inches from the floor).",
    "Suggestions":["Replace to an adjustable height counter"],
    "Sources":[
    {"name":"2010 ADA Standards for Accessible Design", "url":"https://www.ada.gov/regs2010/2010ADAStandards/2010ADAstandards.htm"},
    {"name":"Aging in place: Designing, adapting, and enhancing the home environment","url":"https://scholar.google.com/scholar?hl=en&as_sdt=0%2C48&q=Aging+in+Place+Designing%2C+Adapting%2C+and+Enhancing+the+Home+Environment&btnG="}]
    }
},

"Cabinet": {
    "Pos_Height": {
      "Community": ["Wheelchair"],
      "Dependency": null,
      "Dimension": {
        "Comparison": "LessThan",
        "Value": [27]
      },
      "RelativePosition":{
            "Comparison":null,
            "Value":null
        },
  "Existence": null,
  "Note": null,
  "Message": "Warning: The cabinet is too TALL!",
  "Description": "According to ADA compliance, the height of cabinets should be no more than 27 inches from the floor.",
  "Suggestions": [ "Move things you frequently use to places within easy reach."],
  "Sources": [
    {"name":"2010 ADA Standards for Accessible Design", "url":"https://www.ada.gov/regs2010/2010ADAStandards/2010ADAstandards.htm"}, {"name":"HSSAT","url":"https://www.tompkinscountyny.gov/files2/cofa/documents/hssat_v3.pdf"}]
    }
  },
  
"GrabBar_Existence_Tub": {
"ExistenceOrNot": {
  "Community": ["Wheelchair", "Elder"],
  "Dependency": ["Tub"],
  "Dimension": {
    "Comparison": null,
    "Value": null
  },
  "RelativePosition":{
            "Comparison":"LessThan",
            "Value":[27]
        },
  "Existence": true,
  "Note": null,
  "Message": "Warning: No grab bar detected near tub!",
  "Description": "For safety, there should be grab bars near tub.",
  "Suggestions": ["Add a bath grab bar on the wall or a clamp-on grab bar to the tub."],
  "Sources": [
    {"name":"HSSAT","url":"https://www.tompkinscountyny.gov/files2/cofa/documents/hssat_v3.pdf"}]
}
},
"Knives": {
"ExistenceOrNot": {
  "Community": ["Children"],
  "Dependency": null,
  "Dimension": {
    "Comparison": null,
    "Value": null
  },
  "RelativePosition":{
            "Comparison":null,
            "Value":null
        },
  "Existence": false,
  "Note": null,
  "Message": "Warning: Knives have been detected in a dangerous place!",
  "Description": "For safety, no knives should be present on reachable surface.",
  "Suggestions": ["Move out of reach of children"],
  "Sources": []
}
\end{minted}
\onecolumn
\section{All supported accessibility and safety issues}
\label{appendix:rubrics}
\begin{table*}[!ht]
\caption{20-item RASSAR Accessibility and Safety Issues.}
  \label{tab:allissues}
\resizebox{0.8\textwidth}{!}{\begin{tabular}{llll}
\hline
\textbf{Issue name}          & \textbf{Category}      & \textbf{Rubirc (inch)}         & \textbf{Rubric (cm)}             \\ \hline
Bed height                   & Object Dimension       & 20 - 23                        & 50.8 - 58.42                     \\
\rowcolor[HTML]{EFEFEF} 
Table height                 & Object Dimension       & 28 - 34                        & 71.1 - 86.4                      \\
Counter height               & Object Dimension       & 28 - 34                        & 71.1 - 86.4                      \\
\rowcolor[HTML]{EFEFEF} 
Door Radius                  & Object Dimension       & \textgreater 32 & \textgreater 81.3 \\
Opening Width                & Object Dimension       & \textgreater 32 & \textgreater 81.3 \\
\rowcolor[HTML]{EFEFEF} 
Cabinet height               & Object Position        & \textless 27    & \textless 68.6    \\
Sink Height                  & Object Position        & \textless 17    & \textless 43.1    \\
\rowcolor[HTML]{EFEFEF} 
Door Handle Height           & Object Position        & 34 - 48                        & 86.4 - 122                       \\
Knob Height                  & Object Position        & 34 - 48                        & 86.4 - 122                       \\
\rowcolor[HTML]{EFEFEF} 
Light switch                 & Object Position        & 15 - 48                        & 38.1 - 122                       \\
Electric socket              & Object Position        & 15 - 48                        & 38.1 - 122                       \\
\rowcolor[HTML]{EFEFEF} 
Grab bar height for adults   & Object Position        & 33 - 36                        & 83.8 - 91.4                      \\
Grab bar height for children & Object Position        & 18 - 27                        & 45.7 - 68.6                      \\
\rowcolor[HTML]{EFEFEF} 
Fire alarm                   & Lack of Assistive Item & \multicolumn{2}{c}{\cellcolor[HTML]{EFEFEF}Should exist}          \\
Grab bar near tub            & Lack of Assistive Item & \multicolumn{2}{c}{Should exist}                                  \\
\rowcolor[HTML]{EFEFEF} 
Grab bar near toilet         & Lack of Assistive Item & \multicolumn{2}{c}{\cellcolor[HTML]{EFEFEF}Should exist}          \\
Rug                          & Risky Item             & \multicolumn{2}{c}{Shouldn't exist}                               \\
\rowcolor[HTML]{EFEFEF} 
Scissors                     & Risky Item             & \multicolumn{2}{c}{\cellcolor[HTML]{EFEFEF}Shouldn't exist}       \\
Knife                        & Risky Item             & \multicolumn{2}{c}{Shouldn't exist}                               \\
\rowcolor[HTML]{EFEFEF} 
Medication                   & Risky Item             & \multicolumn{2}{c}{\cellcolor[HTML]{EFEFEF}Shouldn't exist}       \\ \hline
\end{tabular}}
\end{table*}
\newpage
\section{Technical Evaluation Full Results}
\label{appendix:eval}

\begin{table*}[h!]
\centering
    \caption{Information of ten scanned home spaces and RASSAR evaluation results on them. GT refers to ground truth obtained from the RASSAR accessibility rubrics. Scanning time was measured in seconds. \rev{ Definition of precision, recall, accuracy, F1 score and Krippendorff's Alpha are introduced in \autoref{subsec:evalresults}}}
     \label{tab:evaluationresultsFULL}
\begin{tabular}{lrlrr|rrrrrr}
\hline
\textbf{Space}                                & \multicolumn{1}{l}{\textbf{\begin{tabular}[c]{@{}l@{}}Size \\ (sqm)\end{tabular}}} & \textbf{\begin{tabular}[c]{@{}l@{}}Home\\ Type\end{tabular}} & \multicolumn{1}{l}{\textbf{\begin{tabular}[c]{@{}l@{}}Rooms\\ Scanned\end{tabular}}} & \multicolumn{1}{l|}{\textbf{\begin{tabular}[c]{@{}l@{}}Count \\ of GT\end{tabular}}} & \multicolumn{1}{l}{\textbf{Prec.}} & \multicolumn{1}{l}{\textbf{Recall}} & \multicolumn{1}{l}{\textbf{\begin{tabular}[c]{@{}l@{}}F1\\ Score\end{tabular}}} & \multicolumn{1}{l}{\textbf{Accuracy}} & \multicolumn{1}{l}{\textbf{\begin{tabular}[c]{@{}l@{}}Krippendorf \\ Alpha\end{tabular}}} & \multicolumn{1}{l}{\textbf{\begin{tabular}[c]{@{}l@{}}Scan \\ Time\end{tabular}}} \\ \hline
                                              &                                                                                    &                                                              &                                                                                      &                                                                                      & 0.74                               & 0.93                                & 0.82                                                                            & 0.70                                  &                                                                                           &                                                                                   \\
                                              &                                                                                    &                                                              &                                                                                      &                                                                                      & 0.72                               & 0.87                                & 0.79                                                                            & 0.65                                  &                                                                                           &                                                                                   \\
\multirow{-3}{*}{S1}                          & \multirow{-3}{*}{65}                                                               & \multirow{-3}{*}{Apt}                                        & \multirow{-3}{*}{3}                                                                  & \multirow{-3}{*}{15}                                                                 & 0.78                               & 0.93                                & 0.85                                                                            & 0.74                                  & \multirow{-3}{*}{0.73}                                                                    & \multirow{-3}{*}{113}                                                             \\
\rowcolor[HTML]{EFEFEF} 
\cellcolor[HTML]{EFEFEF}                      & \cellcolor[HTML]{EFEFEF}                                                           & \cellcolor[HTML]{EFEFEF}                                     & \cellcolor[HTML]{EFEFEF}                                                             & \cellcolor[HTML]{EFEFEF}                                                             & 0.64                               & 0.82                                & 0.72                                                                            & 0.56                                  & \cellcolor[HTML]{EFEFEF}                                                                  & \cellcolor[HTML]{EFEFEF}                                                          \\
\rowcolor[HTML]{EFEFEF} 
\cellcolor[HTML]{EFEFEF}                      & \cellcolor[HTML]{EFEFEF}                                                           & \cellcolor[HTML]{EFEFEF}                                     & \cellcolor[HTML]{EFEFEF}                                                             & \cellcolor[HTML]{EFEFEF}                                                             & 0.73                               & 0.73                                & 0.73                                                                            & 0.57                                  & \cellcolor[HTML]{EFEFEF}                                                                  & \cellcolor[HTML]{EFEFEF}                                                          \\
\rowcolor[HTML]{EFEFEF} 
\multirow{-3}{*}{\cellcolor[HTML]{EFEFEF}S2}  & \multirow{-3}{*}{\cellcolor[HTML]{EFEFEF}63}                                       & \multirow{-3}{*}{\cellcolor[HTML]{EFEFEF}Apt}                & \multirow{-3}{*}{\cellcolor[HTML]{EFEFEF}2}                                          & \multirow{-3}{*}{\cellcolor[HTML]{EFEFEF}11}                                         & 0.78                               & 0.64                                & 0.70                                                                            & 0.54                                  & \multirow{-3}{*}{\cellcolor[HTML]{EFEFEF}0.7}                                             & \multirow{-3}{*}{\cellcolor[HTML]{EFEFEF}120}                                     \\
                                              &                                                                                    &                                                              &                                                                                      &                                                                                      & 0.78                               & 0.88                                & 0.82                                                                            & 0.70                                  &                                                                                           &                                                                                   \\
                                              &                                                                                    &                                                              &                                                                                      &                                                                                      & 0.82                               & 0.75                                & 0.78                                                                            & 0.64                                  &                                                                                           &                                                                                   \\
\multirow{-3}{*}{S3}                          & \multirow{-3}{*}{45}                                                               & \multirow{-3}{*}{House}                                      & \multirow{-3}{*}{4}                                                                  & \multirow{-3}{*}{24}                                                                 & 0.95                               & 0.75                                & 0.84                                                                            & 0.72                                  & \multirow{-3}{*}{0.67}                                                                    & \multirow{-3}{*}{148}                                                             \\
\rowcolor[HTML]{EFEFEF} 
\cellcolor[HTML]{EFEFEF}                      & \cellcolor[HTML]{EFEFEF}                                                           & \cellcolor[HTML]{EFEFEF}                                     & \cellcolor[HTML]{EFEFEF}                                                             & \cellcolor[HTML]{EFEFEF}                                                             & 0.89                               & 0.73                                & 0.80                                                                            & 0.67                                  & \cellcolor[HTML]{EFEFEF}                                                                  & \cellcolor[HTML]{EFEFEF}                                                          \\
\rowcolor[HTML]{EFEFEF} 
\cellcolor[HTML]{EFEFEF}                      & \cellcolor[HTML]{EFEFEF}                                                           & \cellcolor[HTML]{EFEFEF}                                     & \cellcolor[HTML]{EFEFEF}                                                             & \cellcolor[HTML]{EFEFEF}                                                             & 1.00                               & 0.82                                & 0.90                                                                            & 0.82                                  & \cellcolor[HTML]{EFEFEF}                                                                  & \cellcolor[HTML]{EFEFEF}                                                          \\
\rowcolor[HTML]{EFEFEF} 
\multirow{-3}{*}{\cellcolor[HTML]{EFEFEF}S4}  & \multirow{-3}{*}{\cellcolor[HTML]{EFEFEF}55}                                       & \multirow{-3}{*}{\cellcolor[HTML]{EFEFEF}Apt}                & \multirow{-3}{*}{\cellcolor[HTML]{EFEFEF}3}                                          & \multirow{-3}{*}{\cellcolor[HTML]{EFEFEF}11}                                         & 0.82                               & 0.82                                & 0.82                                                                            & 0.69                                  & \multirow{-3}{*}{\cellcolor[HTML]{EFEFEF}0.82}                                            & \multirow{-3}{*}{\cellcolor[HTML]{EFEFEF}80}                                      \\
                                              &                                                                                    &                                                              &                                                                                      &                                                                                      & 0.91                               & 0.91                                & 0.91                                                                            & 0.83                                  &                                                                                           &                                                                                   \\
                                              &                                                                                    &                                                              &                                                                                      &                                                                                      & 0.91                               & 0.91                                & 0.91                                                                            & 0.83                                  &                                                                                           &                                                                                   \\
\multirow{-3}{*}{S5}                          & \multirow{-3}{*}{50}                                                               & \multirow{-3}{*}{Apt}                                        & \multirow{-3}{*}{3}                                                                  & \multirow{-3}{*}{11}                                                                 & 1.00                               & 0.91                                & 0.95                                                                            & 0.91                                  & \multirow{-3}{*}{1}                                                                       & \multirow{-3}{*}{84}                                                              \\
\rowcolor[HTML]{EFEFEF} 
\cellcolor[HTML]{EFEFEF}                      & \cellcolor[HTML]{EFEFEF}                                                           & \cellcolor[HTML]{EFEFEF}                                     & \cellcolor[HTML]{EFEFEF}                                                             & \cellcolor[HTML]{EFEFEF}                                                             & 0.92                               & 0.85                                & 0.88                                                                            & 0.79                                  & \cellcolor[HTML]{EFEFEF}                                                                  & \cellcolor[HTML]{EFEFEF}                                                          \\
\rowcolor[HTML]{EFEFEF} 
\cellcolor[HTML]{EFEFEF}                      & \cellcolor[HTML]{EFEFEF}                                                           & \cellcolor[HTML]{EFEFEF}                                     & \cellcolor[HTML]{EFEFEF}                                                             & \cellcolor[HTML]{EFEFEF}                                                             & 0.85                               & 0.85                                & 0.85                                                                            & 0.73                                  & \cellcolor[HTML]{EFEFEF}                                                                  & \cellcolor[HTML]{EFEFEF}                                                          \\
\rowcolor[HTML]{EFEFEF} 
\multirow{-3}{*}{\cellcolor[HTML]{EFEFEF}S6}  & \multirow{-3}{*}{\cellcolor[HTML]{EFEFEF}90}                                       & \multirow{-3}{*}{\cellcolor[HTML]{EFEFEF}Apt}                & \multirow{-3}{*}{\cellcolor[HTML]{EFEFEF}4}                                          & \multirow{-3}{*}{\cellcolor[HTML]{EFEFEF}13}                                         & 1.00                               & 0.77                                & 0.87                                                                            & 0.77                                  & \multirow{-3}{*}{\cellcolor[HTML]{EFEFEF}0.83}                                            & \multirow{-3}{*}{\cellcolor[HTML]{EFEFEF}125}                                     \\
                                              &                                                                                    &                                                              &                                                                                      &                                                                                      & 0.76                               & 0.87                                & 0.81                                                                            & 0.68                                  &                                                                                           &                                                                                   \\
                                              &                                                                                    &                                                              &                                                                                      &                                                                                      & 0.81                               & 0.87                                & 0.84                                                                            & 0.72                                  &                                                                                           &                                                                                   \\
\multirow{-3}{*}{S7}                          & \multirow{-3}{*}{65}                                                               & \multirow{-3}{*}{Apt}                                        & \multirow{-3}{*}{3}                                                                  & \multirow{-3}{*}{15}                                                                 & 0.93                               & 0.87                                & 0.90                                                                            & 0.81                                  & \multirow{-3}{*}{0.62}                                                                    & \multirow{-3}{*}{96}                                                              \\
\rowcolor[HTML]{EFEFEF} 
\cellcolor[HTML]{EFEFEF}                      & \cellcolor[HTML]{EFEFEF}                                                           & \cellcolor[HTML]{EFEFEF}                                     & \cellcolor[HTML]{EFEFEF}                                                             & \cellcolor[HTML]{EFEFEF}                                                             & 1.00                               & 0.70                                & 0.82                                                                            & 0.70                                  & \cellcolor[HTML]{EFEFEF}                                                                  & \cellcolor[HTML]{EFEFEF}                                                          \\
\rowcolor[HTML]{EFEFEF} 
\cellcolor[HTML]{EFEFEF}                      & \cellcolor[HTML]{EFEFEF}                                                           & \cellcolor[HTML]{EFEFEF}                                     & \cellcolor[HTML]{EFEFEF}                                                             & \cellcolor[HTML]{EFEFEF}                                                             & 1.00                               & 0.60                                & 0.75                                                                            & 0.60                                  & \cellcolor[HTML]{EFEFEF}                                                                  & \cellcolor[HTML]{EFEFEF}                                                          \\
\rowcolor[HTML]{EFEFEF} 
\multirow{-3}{*}{\cellcolor[HTML]{EFEFEF}S8}  & \multirow{-3}{*}{\cellcolor[HTML]{EFEFEF}50}                                       & \multirow{-3}{*}{\cellcolor[HTML]{EFEFEF}Apt}                & \multirow{-3}{*}{\cellcolor[HTML]{EFEFEF}3}                                          & \multirow{-3}{*}{\cellcolor[HTML]{EFEFEF}10}                                         & 1.00                               & 0.80                                & 0.89                                                                            & 0.80                                  & \multirow{-3}{*}{\cellcolor[HTML]{EFEFEF}0.69}                                            & \multirow{-3}{*}{\cellcolor[HTML]{EFEFEF}80}                                      \\
                                              &                                                                                    &                                                              &                                                                                      &                                                                                      & 0.78                               & 1.00                                & 0.88                                                                            & 0.78                                  &                                                                                           &                                                                                   \\
                                              &                                                                                    &                                                              &                                                                                      &                                                                                      & 0.75                               & 0.86                                & 0.80                                                                            & 0.67                                  &                                                                                           &                                                                                   \\
\multirow{-3}{*}{S9}                          & \multirow{-3}{*}{24}                                                               & \multirow{-3}{*}{House}                                      & \multirow{-3}{*}{2}                                                                  & \multirow{-3}{*}{7}                                                                  & 0.67                               & 0.86                                & 0.75                                                                            & 0.60                                  & \multirow{-3}{*}{-0.05}                                                                   & \multirow{-3}{*}{53}                                                              \\
\rowcolor[HTML]{EFEFEF} 
\cellcolor[HTML]{EFEFEF}                      & \cellcolor[HTML]{EFEFEF}                                                           & \cellcolor[HTML]{EFEFEF}                                     & \cellcolor[HTML]{EFEFEF}                                                             & \cellcolor[HTML]{EFEFEF}                                                             & 0.92                               & 0.86                                & 0.89                                                                            & 0.80                                  & \cellcolor[HTML]{EFEFEF}                                                                  & \cellcolor[HTML]{EFEFEF}                                                          \\
\rowcolor[HTML]{EFEFEF} 
\cellcolor[HTML]{EFEFEF}                      & \cellcolor[HTML]{EFEFEF}                                                           & \cellcolor[HTML]{EFEFEF}                                     & \cellcolor[HTML]{EFEFEF}                                                             & \cellcolor[HTML]{EFEFEF}                                                             & 0.92                               & 0.86                                & 0.89                                                                            & 0.80                                  & \cellcolor[HTML]{EFEFEF}                                                                  & \cellcolor[HTML]{EFEFEF}                                                          \\
\rowcolor[HTML]{EFEFEF} 
\multirow{-3}{*}{\cellcolor[HTML]{EFEFEF}S10} & \multirow{-3}{*}{\cellcolor[HTML]{EFEFEF}60}                                       & \multirow{-3}{*}{\cellcolor[HTML]{EFEFEF}House}              & \multirow{-3}{*}{\cellcolor[HTML]{EFEFEF}3}                                          & \multirow{-3}{*}{\cellcolor[HTML]{EFEFEF}14}                                         & 0.93                               & 0.93                                & 0.93                                                                            & 0.87                                  & \multirow{-3}{*}{\cellcolor[HTML]{EFEFEF}0.43}                                            & \multirow{-3}{*}{\cellcolor[HTML]{EFEFEF}100}                                     \\ \hline
\multicolumn{5}{r|}{\textbf{Average}}                                                                                                                                                                                                                                                                                                                                           & \textbf{0.86}                      & \textbf{0.83}                       & \textbf{0.84}                                                                   & \textbf{0.72}                         & \textbf{0.64}                                                                             & \textbf{99.9}                                                                     \\ \hline
\end{tabular}
\end{table*}

\end{document}